\documentclass[sigconf,screen,nonacm]{acmart}

\usepackage{booktabs} 
\usepackage{subfigure}
\usepackage{multirow}
\usepackage[ruled]{algorithm2e}

\DeclareMathOperator{\E}{\mathbb{E}}

\begin{document}
\title{GAN-based Matrix Factorization for Recommender Systems}
\titlenote{Accepted as a long paper at the 37\textsuperscript{th} ACM/SIGAPP Symposium on Applied Computing (SAC '22), Recommender Systems track, \url{https://doi.org/10.1145/3477314.3507099}.}

\author{Ervin Dervishaj}
\affiliation{
    \institution{ContentWise, Politecnico di Milano}
    \city{Milan}
    \country{Italy}
}
\email{ervin.dervishaj@mail.polimi.it}

\author{Paolo Cremonesi}
\affiliation{
    \institution{Politecnico di Milano}
    \city{Milan}
    \country{Italy}
}
\email{paolo.cremonesi@polimi.it}

\begin{abstract}
Proposed in 2014, Generative Adversarial Networks (GAN) initiated a fresh interest in generative modelling. They immediately achieved state-of-the-art in image synthesis, image-to-image translation, text-to-image generation, image inpainting and have been used in sciences ranging from medicine to high-energy particle physics. Despite their popularity and ability to learn arbitrary distributions, GAN have not been widely applied in recommender systems (RS). Moreover, only few of the techniques that have introduced GAN in RS have employed them directly as a collaborative filtering (CF) model.
  
In this work we propose a new GAN-based approach that learns user and item latent factors in a matrix factorization setting for the generic top-N recommendation problem. Following the vector-wise GAN training approach for RS introduced by CFGAN, we identify 2 unique issues when utilizing GAN for CF. We propose solutions for both of them by using an autoencoder as discriminator and incorporating an additional loss function for the generator. We evaluate our model, GANMF, through well-known datasets in the RS community and show improvements over traditional CF approaches and GAN-based models. Through an ablation study on the components of GANMF we aim to understand the effects of our architectural choices. Finally, we provide a qualitative evaluation of the matrix factorization performance of GANMF.
\end{abstract}



\settopmatter{printfolios=true}
\maketitle

\section{Introduction}
\label{intro}
With the ever-increasing amount of available digital information, recommender systems (RS) are essential tools in filtering out the content presented to users, providing a personalized experience in various domains. The data generated through the interaction of users with RS gave rise to \emph{collaborative filtering} \cite{ricci2011introduction} (CF) which utilizes such user-item interactions to create models that can provide high quality recommendations. Within CF, \emph{latent factor} models are a family of mathematical models that project both users and items into a latent space. Matrix factorization (MF) is the most successful latent factor model for CF \cite{koren2009matrix,funk2006netflix}, made famous during the Netflix Prize challenge. MF expresses the preference of a user on an item as the dot product between their latent factors. 

Generative Adversarial Networks (GAN) \cite{goodfellow2014generative}, proposed by Goodfellow et al. in 2014, are generative models that use neural networks to learn arbitrary probability distributions. GAN have shown impressive results in estimating high-dimensional distributions in computer vision \cite{brock2018large,isola2017image,zhang2017stackgan}, natural language processing \cite{yu2017seqgan,lin2017adversarial} and various other fields like physics \cite{de2017learning,paganini2018calogan} and medicine \cite{schlegl2019f}. They are an active area of research and multiple architectural variants have been proposed like conditional GAN \cite{mirza2014conditional}, EBGAN \cite{zhao2016energy}, InfoGAN \cite{chen2016infogan}, etc. Despite their popularity, GAN have not seen wide adoption in RS. The first work to incorporate GAN in the context of RS was IRGAN \cite{wang2017irgan}, proposed in 2017, albeit focusing more generally in merging discriminative and generative paradigms in information retrieval (IR). GraphGAN \cite{wang2018graphgan} exploits the graph structure of user-item interactions and utilizes GAN to learn embeddings for nodes in this graph. CFGAN \cite{chae2018cfgan} identifies a problem with IRGAN's training and proposes \emph{vector-wise} training for GAN in RS. Other works utilize GAN with additional information beside the user-item interactions \cite{ren2020sequential,cai2018generative}, to alleviate the sparsity of RS ratings \cite{chae2019rating,wang2019enhancing} and to model contextual recommendations \cite{bharadhwaj2018recgan,liu2019geo}. 

Since the adoption of GAN in RS is still in its early phases, we believe there is still room for improvement, especially in using GAN \textbf{explicitly as a CF model}. Current GAN approaches like CFGAN attempt to generate the preferences of a specific user on all items by learning from the real preferences of the user. However, RS are characterized by a high number of items and \emph{a single set} of preferences per user, which makes generating user-specific preferences non-trivial. Based on this, our main contributions in this work are: 
\begin{itemize}
    \item We identify the two issues mentioned above when employing GAN for CF. Motivated by them, we derive GANMF, a new conditional GAN-based latent factor model aimed at the generic top-N recommendation problem under implicit feedback.
    \item We show that GANMF outperforms both traditional and other GAN baselines in two ranking metrics on 3 well-known datasets in RS community.
    \item We perform an ablation study on the components of GANMF to better understand how they affect the model performance.
    \item We investigate the MF model learned by GANMF in terms of the number of latent factors and users with fewer interactions.
\end{itemize}

The rest of this paper is structured as follows. Section \ref{preliminaries} provides a short overview of GAN followed by a presentation of other GAN-based RS. In section \ref{motivation}, we motivate our work through the two key issues of GAN in CF. Section \ref{ganmf} details our proposed model and its components. In section \ref{experiments}, we provide the settings of our experiments (section \ref{settings_evaluation}) and discuss the obtained results (section \ref{baselines}). We present an ablation study in section \ref{ablation_study} and further examine our model in the context of MF in section \ref{sec:mf_learned}. Finally, we conclude with section \ref{conclusion}.

\section{Preliminaries}
\label{preliminaries}
We formally present the generic problem of top-N recommendation. Given a set of users $U$, a set of items $I$ and users' past feedback on these items, top-N recommendation is the problem of recommending to every user $u \in U$ a subset of items from $I$ that $u$ (has not previously interacted with) is more likely to enjoy. We can arrange the past feedback into a user rating matrix (URM) of shape $|U| \times |I|$. Every cell $(u, i)$ of the URM represents the feedback of user $u$ on item $i$. In this work we focus only on implicit feedback, given that it is more abundant and easier to secure \cite{koren2009matrix}. In this case, cell $(u,i)$ of URM has a value of 1 if user $u$ has shown interest in item $i$ or 0 otherwise. Each row of the URM represents the historical profile of a user whereas each column the historical profile of an item.

\subsection{Generative Adversarial Networks}
GAN are part of a class of generative models called implicit density estimating generative models \cite{goodfellow2016nips}. They do not assume a fixed form of the data distribution but build a model for the distribution from which we can sample. In a GAN, 2 players are pitted against one another in a minimax zero-sum game. One of the players is the \textbf{generator} ($\mathcal{G}$) and the other the \textbf{discriminator} ($\mathcal{D}$). $\mathcal{G}$, takes as input a noise vector \textbf{z} sampled from a predefined distribution $p_{\mathbf{z}}$ and outputs a synthetic data point in the real data space. $\mathcal{D}$ on the other hand, takes as input data coming from the real training data and from the generator and is tasked to differentiate the source of its input. The generator is trained so that it can fool the discriminator into classifying the data it generates as real data. In this setup, both the discriminator and generator optimize the same objective function:
\begin{equation}
    \min_{\mathcal{G}} \max_{\mathcal{D}} \E_{\mathbf{x}\sim{p_{data}}} [log \, D(\mathbf{x})] + \E_{\mathbf{z}\sim{p_{\mathbf{z}}}} \big[log\Big(1-D\big(G(\mathbf{z})\big)\Big)\big]
\end{equation}
Goodfellow et al. theoretically prove that for the generator to learn the distribution of the real data, the discriminator must be maximally confused about the source of its input.

\subsubsection{Conditional GAN (cGAN)}
\label{cgan}
This is a GAN variant in which the generator is guided to produce data that belongs to a given class \cite{mirza2014conditional}. This is achieved by concatenating the class on which we want to condition the generation process to the input of both discriminator and generator. The objective function in a cGAN is thus changed to:
\begin{equation}
    \min_{\mathcal{G}} \max_{\mathcal{D}} \E_{\mathbf{x}\sim{p_{data}}} [log \, D(\mathbf{x}|c)] + \E_{\mathbf{z}\sim{p_{\mathbf{z}}}} \big[log\Big(1-D\big(G(\mathbf{z}|c)\big)\Big)\big]
\end{equation}

\subsection{Related Work}
\label{sec:related_work}
IRGAN pioneered GAN in IR and RS. It focuses in providing a unification of generative and discriminative paradigms of modelling in IR. Given a set of queries \{$q_{1}, \ldots, q_{N}$\}, a set of documents \{$d_{1}, \ldots, d_{M}$\} and the true relevancy distribution $p_{\text{true}}(d|q,r)$ of documents to queries, IRGAN learns $p_{\theta}(d|q,r)$ through a generator $G$. $p_{\theta}(d|q,r)$ is such that, when sampled from, a binary classifier $D$ cannot distinguish whether the document is coming from $p_{\theta}$ or from $p_{\text{true}}$. IRGAN optimizes the following minimax function:
\begin{equation}
    \footnotesize
    \label{eq:irgan_binary_func}
        \min_{\theta} \max_{\phi} \sum_{n=1}^N\Big(
        \E_{d \sim{p_{\text{true}}(d|q_{n},r)}}[log \, D(d|q_{n})] \: + \:
        \E_{d \sim p_{\theta}(d|q_{n},r)}\big[log\big(1-D(d|q_{n})\big)\big]\Big)
\end{equation}
where the parameters $\theta$ of $G$ are updated through the REINFORCE algorithm \cite{williams1992simple} due to the non-differentiable discrete sampling from $p_{\theta}$. CFGAN \cite{chae2018cfgan}, a cGAN approach for RS, takes the training process of IRGAN and experimentally shows that the discrete sampling operation in the generator deteriorates the performance of the discriminator. It proposes \emph{vector-wise} training as a solution where the generator produces full user historical profiles. The discriminator differentiates between generated profiles and real user profiles coming from the URM. CAAE \cite{chae2019collaborative} uses 2 autoencoder-based generators as a replacement for MF-based generator in IRGAN and pairs those with BPR \cite{rendle2012bpr} loss in the discriminator. To feed the discriminator, one of the generators samples positive items for a given user whereas the other samples negative ones. $\text{RAGAN}$ \cite{chae2019rating} and AugCF \cite{wang2019enhancing} focus on alleviating the sparsity of the URM and then apply traditional CF models to provide recommendations. $\text{RAGAN}$ utilizes the recommendation capabilities of CFGAN to generate explicit ratings for items and uses them to impute the missing ratings in the URM. $\text{RAGAN}^{\text{BT}}$ \cite{chae2019rating} fixes $\text{RAGAN}$'s bias towards generating \emph{high-value ratings} by incorporating the application of CDAE \cite{wu2016collaborative} on the user historical profile before using CFGAN to fill the URM. AugCF is a cGAN where the generator takes as input a user, unvisited items for this user, associated side information and a class (like or dislike) and outputs a plausible item for the user under the selected class. The discriminator in AugCF takes two roles; in a first phase it acts as a classifier whereas in a second phase it functions as a pure CF model. Other works use GAN with additional information beside the user-item interactions. AugCF can also be considered such a model. \cite{cai2018generative} tackles the task of learning node representations in a bibliographic network by combining the content of papers with the adjacency matrix between papers and authors.

\section{Motivation}
\label{motivation}
We highlight two issues that arise when utilizing a cGAN to \textbf{generate plausible user profiles} (as in the case of CFGAN). The discriminator of a cGAN is usually a binary classifier; it outputs a scalar value indicating the probability that the input is coming from the real data. 
However, the output of the generator in the case of CFGAN is very high dimensional; specifically its dimension is the length of a user historical profile $|I|$, which for some datasets can be in hundreds of thousands or even millions of elements.
Updating the weights of the generator through the gradient of a single scalar value in the discriminator output poses difficulties for learning the generator function. As an example, given user $u$'s real profile $I^{*}_{u}$, we consider the case of 2 generated profiles for user $u$, $\hat{I}^1_{u}$ and $\hat{I}^2_{u}$. $\hat{I}^1_{u}$ differs from $I^{*}_{u}$ on only some items whereas $\hat{I}^2_{u}$ is the inverse of $I^{*}_{u}$. 
For an optimal binary classifier, both $\hat{I}^1_{u}$ and $\hat{I}^2_{u}$ are fake and the gradient of the loss propagated back to the generator would be more or less the same for both. However, under some distance metric, we have:
\begin{equation}
    \big\|I^{*}_{u} - \hat{I}^{1}_{u}\big\| \leq \big\|I^{*}_{u} - \hat{I}^{2}_{u}\big\|
\end{equation}
yet the gradient coming from the discriminator to the generator would not make this distinction clear.

A cGAN takes a class label as part of the input in both generator and discriminator so that the generated data are conditioned on the class. Applications of cGAN usually involve datasets with multiple classes where for each class there are hundreds or thousands of samples from the training set. However, in the case of CFGAN where a user identifier is considered the conditioning class, there is only a single profile per user. Any model's ability to learn the input-target function from a single data point per target label is very limited.

\begin{figure}[!htbp]
  \centering
  \includegraphics[width=\linewidth]{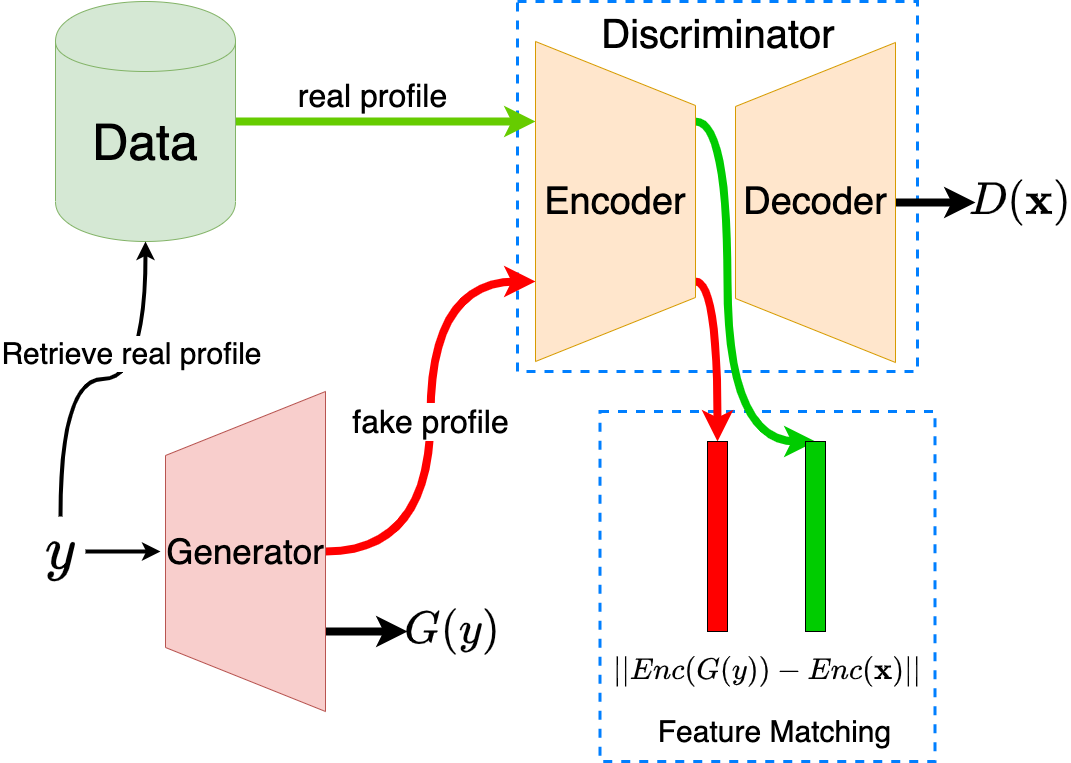}
  \caption{GANMF architecture. The generator is fed a user identifier and produces a plausible user profile. The discriminator is trained with real profiles retrieved from the URM and with fake profiles coming from the generator.}
  \label{fig:ganmf}
\end{figure}

\section{GANMF}
\label{ganmf}
We propose a new model, GANMF, that circumvents both issues raised in section \ref{motivation}. GANMF takes the form of a cGAN; the generator is conditioned through a user identifier $\mathbf{y}$ to produce a plausible user profile belonging to $\mathbf{y}$, whereas the discriminator is trained to differentiate between profiles produced by the generator and real profiles retrieved from the URM. The optimization of these 2 networks is alternated until the discriminator cannot distinguish the profiles produced by the generator from the real ones. Figure \ref{fig:ganmf} gives a visual depiction of the complete architecture of GANMF.

\subsection{GANMF Discriminator}
\label{ganmf_discriminator}
In order to provide richer gradients to the generator, in GANMF we replace the binary classifier discriminator of cGAN with an autoencoder. The autoencoder takes as input a user profile, either from the URM or one generated by the generator, and outputs its reconstruction. EBGAN \cite{zhao2016energy} was first to introduce an autoencoder-based discriminator as an energy function that assigns low energy to data from the training set and high energy to data produced by the generator, thus differentiating the source of the input of the discriminator. In GANMF, the reconstruction error of the autoencoder acts as the energy function:
\begin{equation}
    D(\mathbf{x}) = \big\|Dec\big(Enc(\mathbf{x})\big) - \mathbf{x}\big\|^{2}_{2}
    \label{eq:recon_loss}
\end{equation}
where $Dec(\cdot)$ and $Enc(\cdot)$ are the respective decoder and encoder functions of the autoencoder and $\|\cdot\|_{2}$ is the Euclidean norm. Given a conditioning user identifier $\mathbf{y}$ and its real profile $\mathbf{x}$, the discriminator $\mathcal{D}$ \textbf{minimizes} a hinge loss function:
\begin{equation}
    \mathcal{L}_{\mathcal{D}}(\mathbf{x}, \mathbf{y}) = D(\mathbf{x}) + \big[mD(\mathbf{x}) - D\big(G(\mathbf{y})\big)\big]^{+} + \lambda_{\mathcal{D}} \big\|\Omega_{\mathcal{D}}\big\|^{2}_{2}
    \label{eq:d_loss}
\end{equation}
where $[\cdot]^+ = \max (0, \cdot)$, $D(\cdot)$ is the discriminator function as defined in (\ref{eq:recon_loss}), $G(\cdot)$ is the generator function, $\lambda_{\mathcal{D}}$ is a regularization coefficient and $\Omega_{\mathcal{D}}$ is the set of parameters of $\mathcal{D}$. Different from EBGAN, instead of a positive margin we use a positive margin coefficient $m$ in order to limit the range of its values. Optimizing (\ref{eq:d_loss}) forces the reconstruction error of real profiles towards zero whereas the reconstruction error of generated profiles $m$-times that of the real ones. Note that different from \cite{sedhain2015autorec} where the reconstruction loss of the autoencoder is computed only on past true interactions, in GANMF the discriminator loss is computed over all items. This is because the autoencoder is only used to differentiate the source of the profile and not directly deriving recommendations from it. Finally, we point out that GANMF discriminator does not take a user conditioning vector like CFGAN, CAAE and RAGAN do.

\subsection{GANMF Generator}
\label{ganmf_generator}
Generator $\mathcal{G}$ of GANMF is a conditional generator that takes as input a conditioning attribute $\mathbf{y}$ that is unique for each user. Contrary to the original formulation of cGAN, we drop the noise vector \textbf{z} in the input in order to have deterministic mapping from the conditioning attribute to the generated profile.

\begin{figure}[t]
    \centering
    \includegraphics[width=\linewidth]{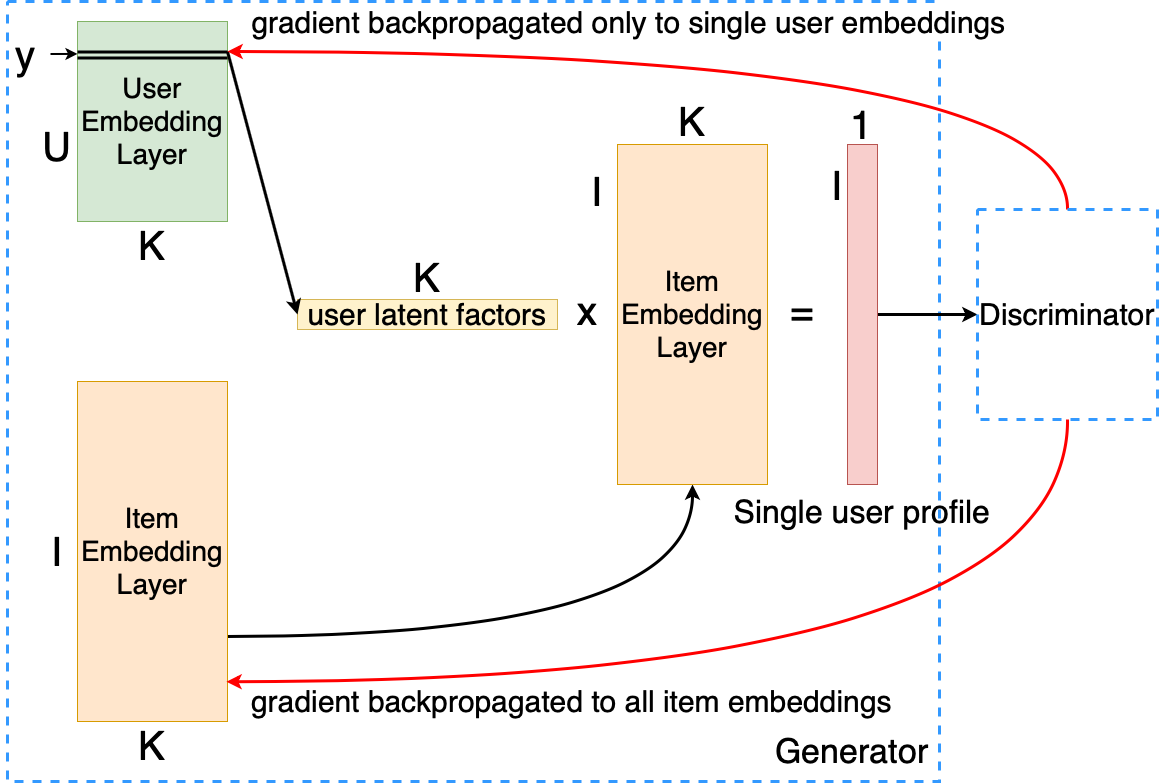}
    \caption{Generator network casted as MF-based approach with embedding layers.}
    \label{fig:mf_embeddings}
\end{figure}

We cast $\mathcal{G}$ into a MF model by using 2 embedding layers, $\Sigma \in \mathbb{R}^{|U| \times K}$ and $V \in \mathbb{R}^{|I| \times K}$, with $\Sigma$ and $V$ being the user and item latent factors and $K$ the number of latent features. Our training data is composed of user-item interactions only, so we use as conditioning attribute for the generator the row number of each user in the URM. In the forward pass of the generator, $\mathbf{y}$ is utilized to retrieve the $\mathbf{y}$-th row from $\Sigma$. A synthetic user profile is produced by (see figure \ref{fig:mf_embeddings}):
\begin{equation}
    G(y) = V \Sigma[\mathbf{y},:]^{\top}
\end{equation}
In order to fool the discriminator, the generator \textbf{minimizes} the reconstruction error of the discriminator on generated profiles:
\begin{equation}
    \mathcal{L}_{\mathcal{G}}(\mathbf{y}) = D\big(G(\mathbf{y})\big) + \lambda_{\mathcal{G}} \big\|\Omega_{\mathcal{G}}\big\|^{2}_{2}
\end{equation}
where $D(\cdot)$ and $G(\cdot)$ are the discriminator and generator functions respectively, $\mathbf{y}$ is the user row in the URM, $\lambda_{\mathcal{G}}$ is the $L_{2}$ regularization coefficient and $\Omega_{\mathcal{G}}$ is the set of parameters of the generator. We note that a clear advantage of using embedding layers is that the number of parameters to be learned by the generator is $\Theta\big(K(|I| + |U|)\big)$, similar to baselines like WRMF \cite{hu2008collaborative}. Moreover, the generator of GANMF is \textbf{simpler} than other GAN-based RS approaches in that it models only the linear interactions between user and item latent factors and does not incorporate non-linearities.

As mentioned in section \ref{motivation}, training GANMF with a single real profile per user (the analogy of a single sample per class) causes the generator $\mathcal{G}$ to disregard the conditioning attribute during the generation process and to produce a single profile for every user which reduces substantially the recommendations' quality.

To alleviate this problem we incorporate in $\mathcal{L}_{\mathcal{G}}$ an additional loss term called \emph{feature matching} \cite{salimans2016improved}. Feature matching is presented as a solution to stabilize the training procedure of GAN by making the generator produce data that match the statistics of the real data. However, in this work we are interested in $\mathcal{G}$ producing user specific profiles. Feature matching is closely related to \emph{Maximum Mean Discrepancy} \cite{gretton2012kernel, tschannen2018recent} which is a distance in probability space between two distributions. Therefore, we modify $\mathcal{L}_{\mathcal{G}}$ by adding this additional loss:
\begin{equation}
    \small
    \label{eq:ganloss_fm}
    \mathcal{L}_{\mathcal{G}}(\mathbf{x},\mathbf{y}) = (1-\alpha) D\big(G(\mathbf{y})\big) + \alpha \Big\|Enc(\mathbf{x}) - Enc\big(G(\mathbf{y})\big)\Big\|^2_{2} + \lambda_{\mathcal{G}} \big\|\Omega_{\mathcal{G}}\big\|^{2}_{2}
\end{equation}
where $Enc(\cdot)$ is the output of the \emph{encoder} in GANMF's discriminator, $\mathbf{y}$ is the user conditioning attribute, $\mathbf{x}$ is $\mathbf{y}$'s real user profile and $\alpha$ is a constant that balances the adversarial and feature matching losses. For GANMF we use a bottlenecked autoencoder which makes the autoencoder learn meaningful features in its \emph{coding} layer \cite{vincent2010stacked}. Minimizing $\mathcal{L}_{\mathcal{G}}$ forces generated profiles to match the distribution of real profiles in the latent space induced by the coding layer. This drives $\mathcal{G}$ to generate profiles that cover the same distribution as the real user profiles in this latent space.

\begin{algorithm}[h]
    \footnotesize
    \label{alg:ganmf}
    \caption{GANMF Training}
    \SetAlgoLined
    \DontPrintSemicolon
    
    \SetKwInOut{Input}{Input}
    \SetKwInOut{Output}{Output}
    
    \SetKwData{U}{U}
    \SetKwData{URM}{URM}
    \SetKwData{Batch}{B}
    \SetKwData{Dpars}{$\Omega_{\mathcal{D}}$}
    \SetKwData{Gpars}{$\Omega_{\mathcal{G}}$}
    \SetKwData{Gen}{$\mathcal{G}$}
    \SetKwData{Dis}{$\mathcal{D}$}
    \SetKwData{Margin}{m}
    \SetKwData{Alpha}{$\alpha$}
    \SetKwData{Dlr}{$\mu_{D}$}
    \SetKwData{Glr}{$\mu_{G}$}
    \SetKwData{NumIters}{numIterations}
    \SetKwData{Iter}{iter}
    \SetKwData{y}{y}
    \SetKwData{fake}{fakeProfiles}
    \SetKwData{real}{realProfiles}
    \SetKwData{Dloss}{$\mathcal{L}_{D}$}
    \SetKwData{Gloss}{$\mathcal{L}_{G}$}
    
    \SetKwFunction{InitVars}{initialize}
    \SetKwFunction{Sample}{sampleBatch}
    \SetKwFunction{Compute}{compute}
    \SetKwFunction{Update}{update}
    \SetKwFunction{D}{$D$}
    \SetKwFunction{G}{$G$}

    \Input{set of users \U, \URM, set of parameters \Dpars, set of parameters \Gpars, margin coefficient \Margin, feature matching coefficient \Alpha, learning rates \Dlr and \Glr, batch size \Batch}
    \Output{trained \Gen model that can generate historical user profiles}
    \BlankLine\BlankLine
    \InitVars{\Dis, \Gen} \;
    \NumIters $\leftarrow$ $\frac{|U|}{|\Batch|}$ \;
    \While{stopping condition not met}{
        \For{\Iter in \NumIters}{
            \tcp{Discriminator learning}
            \y $\leftarrow$ \Sample(\U, \Batch) \;
            \fake $\leftarrow$ \G{\y} \;
            \real $\leftarrow$ \URM[\y,:] \;
            \Dloss $\leftarrow$ \Compute(\real, \fake) \;
            \Dpars $\leftarrow$ \Dpars $-$ \Dlr $\frac{\partial \Dloss}{\partial \Dpars}$ \;
            
            \BlankLine
            
            \tcp{Generator learning}
            \y $\leftarrow$ \Sample(\U, \Batch) \;
            \fake $\leftarrow$ \G{\y} \;
            \Gloss $\leftarrow$ \D(\fake) \;
            \Gpars $\leftarrow$ \Gpars $-$ \Glr $\frac{\partial \Gloss}{\partial \Gpars}$ \;
        }
    }
\end{algorithm}

\section{Experiments}
\label{experiments}

\subsection{Settings and Evaluation}
\label{settings_evaluation}
In our experiments we consider as standard GANMF the model with \textbf{single-hidden-layer autoencoder with linear activations} as discriminator and the generator with \textbf{2 embedding layers} and \textbf{feature matching loss} as described in section \ref{ganmf_generator}. The final recommendations are given by computing each user's profile with the generator and then ranking the items the user has not interacted with. Just like CFGAN, our model also has 2 training modes; user-based (\emph{GANMF-u}) and item-based (\emph{GANMF-i}).

We utilize bayesian optimization\footnote{We use the Python library scikit-optimize (\url{https://scikit-optimize.github.io/stable/)} for the bayesian optimization.} \cite{antenucci2018artist} to find the best hyperparameters for GANMF and the baselines (\textbf{all baselines are trained from scratch in our dataset splits}). For each algorithm we optimize MAP@5 on a holdout set and perform 50 runs with the first 10 being random evaluations that seed the Gaussian Process. For GANMF\footnote{We implemented GANMF with Tensorflow. Both the generator and the discriminator architectures are highly parallelizable through a GPU which helps limit training/inference times.} we use Adam optimizer and tune the following intervals for the hyperparameters\footnote{Epochs and number of latent factors hyperparameters are shared by all baselines. The same interval is used in all models.}:
\begin{itemize}
    \item Number of epochs: a maximum value of 300.
    \item Number of latent factors: integer value in $[1-250]$.
    \item Units in the coding layer of AE: integer value in $[4-1024]$.
    \item Batch size: categorical value in [64, 128, 256, 512, 1024].
    \item Margin coefficient $m$: integer value in $[1-10]$.
    \item Feature matching coefficient $\alpha$: real value in $[0.01-0.5]$ with a uniform prior distribution.
    \item $\mathcal{D}$ and $\mathcal{G}$ learning rates: real value in $[0.0001-0.01]$ with a log-uniform prior distribution.
    \item Regularization coefficient $\lambda_{\mathcal{D}}$: real value in $[10^{-6}-10^{-4}]$ with a log-uniform prior distribution.
    \item Regularization coefficient $\lambda_{\mathcal{G}}$: we set this hyperparameter to 0 since $\mathcal{G}$ does not learn directly from the real training data.
\end{itemize}
The training procedure for GANMF is given by algorithm 1. 

\subsubsection{Datasets}
\label{datasets}
We evaluate GANMF on three well-known datasets in the RS community; MovieLens 1M \cite{harper2015movielens}, MovieLens HetRec \cite{cantador2011second} and LastFM  \cite{cantador2011second}. MovieLens datasets contain explicit user ratings on movies with every user having rated at least 20 movies. LastFM contains music artist listening information in the form of triples \emph{(user, artist, listeningCount)} where \emph{listeningCount} represents how many times the \emph{user} listened to the \emph{artist}. The statistics of the datasets are given on table \ref{tab:dataset_stats}.

\begin{table}[htb]
    \centering
    \caption{Dataset statistics.}
    \begin{tabular}{ccccc}
         \toprule
         \textbf{Dataset} & \textbf{Interactions} & \textbf{Users} & \textbf{Items} & \textbf{Sparsity}\\
         \midrule
         ML 1M & 1000209 & 6040 & 3706 & 95.53\%\\
         ML HetRec & 855598 & 2113 & 10109 & 96.00\% \\ 
         LastFM & 92834 & 1884 & 17626 & 99.72\%\\
         \bottomrule
    \end{tabular}
    \label{tab:dataset_stats}
\end{table}

\begin{figure}[tbh]
    \centering
    \includegraphics[width=\linewidth]{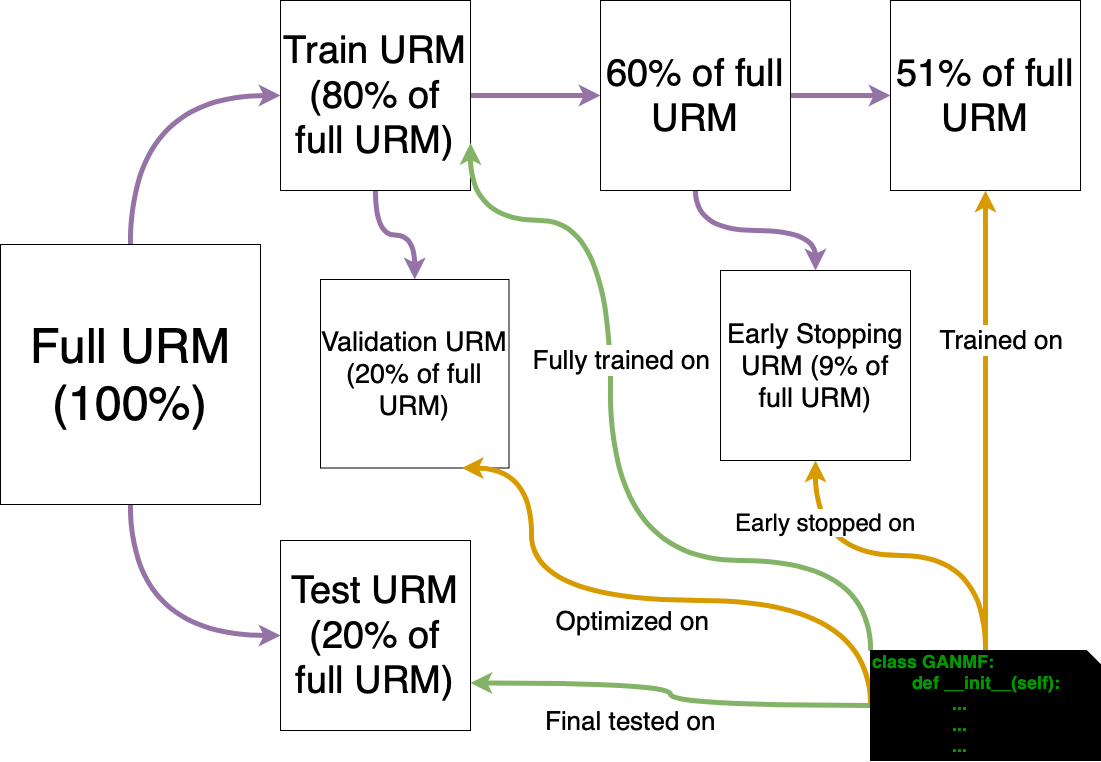}
    \caption{Splitting of the full URM into subsequent sets for each of the operations performed by the algorithms.}
    \label{fig:dataset_splitting}
\end{figure}

In this work we focus on implicit feedback so we drop the movie ratings in MovieLens datasets and \emph{listeningCount} in LastFM and keep only the interactions between users and items from which we build the URM (we denote this as the \emph{full URM}). We randomly split each dataset into train and test sets in 4:1 ratio. In order to have at least one item per user in both sets we consider only users that have interacted with at least 2 items. We use the test set \textbf{only} for the final evaluation of all algorithms. In order to tune hyperparameters we further split the training set to obtain validation and early stopping sets. Once we have the dataset-specific hyperparameters, we fully train each algorithm on the initial train set. In figure \ref{fig:dataset_splitting} we give a summary of the steps from the full URM to each of the sets and how they are used by the algorithms.

\subsubsection{Evaluation}
We compare all algorithms on the test set through 2 different metrics, \emph{normalized discounted cumulative gain} (NDCG) \cite{jarvelin2002cumulated} and \emph{mean average precision} (MAP) on recommendation lists of 5 and 20 items. Both of these metrics take into consideration the ranking of recommended items beside their relevancy. The different cutoffs give us an indication how the algorithms behave with increasing recommendation list length.

\vspace{1em}

The code for all algorithms\footnote{For the baselines and model evaluation we use implementations from \url{https://github.com/MaurizioFD/RecSys_Course_AT_PoliMi}}, along with the datasets' splits, experiments and results can be found in \textbf{\url{https://github.com/edervishaj/GANMF}}.

\subsection{Results}
\label{baselines}
We compare our proposed model with 8 other baseline models aimed at the top-N recommendation problem, including robust baselines \cite{dacrema2019we, dacrema2021troubling}:
\begin{itemize}
    \item \textbf{Top-popular}: non-personalized approach where the most popular items are recommended to every user.
    \item \textbf{PureSVD} \cite{cremonesi2010performance}: MF approach that utilizes SVD to reconstruct the URM. We tune only the number of latent factors for this model.
    \item \textbf{WRMF} \cite{hu2008collaborative}: MF technique that converts implicit feedback into confidence values and employs \emph{alternating least squares} for the computation of user and item latent factors. We tune all the parameters as given in the paper.
    \item \textbf{ItemKNN} \cite{deshpande2004item}: one of the main model-based CF techniques that builds an item-item similarity matrix from the URM. For ItemKNN we tune the similarity\footnote{Considered similarities: dice, jaccard, tversky, asymmetric-cosine, cosine. We report only ItemKNN-$cosine$ since it performed best.}, the neighborhood size and the shrink term.
    \item $\boldsymbol{P^{3}\alpha}$ \cite{cooper2014random}: a graph-based recommendation approach where the similarity between items is expressed as a 2-step random walk starting from users in the bipartite graph of users and items. For this model we tune the neighborhood size and similarity scaling coefficient $\alpha$.
    \item \textbf{SLIM} \cite{ning2011slim}: a machine learning technique that models the item-item similarity matrix and is trained with BPR \cite{rendle2012bpr} loss. For SLIM we tune the hyperparameters as given in the paper.
    \item \textbf{CFGAN} \cite{chae2018cfgan}: GAN-based RS that proposes \emph{vector-wise} training for GAN in RS. We tune all the hyperparameters in the ranges provided by the authors.
    \item \textbf{CAAE} \cite{chae2019collaborative}: GAN-based RS that incorporates BPR loss in the discriminator and uses 2 autoencoder-based generators. For CAAE we tune the hyperparameters in the ranges provided by the authors.
\end{itemize}
For CFGAN and GANMF we give both user and item-based variants. For each of the baselines we perform hyperparameter tuning as explained in section \ref{settings_evaluation}. Table \ref{tab:baselines} summarizes the comparison between GANMF and all baselines.

\begin{table*}[htbp]
    \centering
    \caption{Experimental results of GANMF and chosen baselines over MovieLens 1M, MovieLens HetRec and LastFM. The best results per dataset and per metric are given in bold, second best results are underlined.}
    \normalsize
    \begin{tabular}{c|cccc|cccc|cccc}
        \toprule
        \multirow{3}{*}{Algorithm}
        & \multicolumn{4}{|c}{\textbf{ML 1M}} & \multicolumn{4}{|c}{\textbf{ML HetRec}} & \multicolumn{4}{|c}{\textbf{LastFM}}\\
        & \textbf{NDCG} & \textbf{MAP} & \textbf{NDCG} & \textbf{MAP} & \textbf{NDCG} & \textbf{MAP} & \textbf{NDCG} & \textbf{MAP} & \textbf{NDCG} & \textbf{MAP} & \textbf{NDCG} & \textbf{MAP}\\
        & @5 & @5 & @20 & @20 & @5 & @5 & @20 & @20 & @5 & @5 & @20 & @20\\
        \midrule
        Top Popular & 0.2248 & 0.1544 & 0.1952 & 0.0919 & 0.4770 & 0.3899 & 0.3905 & 0.2475 & 0.0888 & 0.0524 & 0.0947 & 0.0392\\
        PureSVD & 0.4197 & 0.3243 & 0.3644 & 0.2139 & 0.5933 & 0.5126 & 0.5020 & 0.3604 & 0.2263 & 0.1505 & 0.2145 & 0.1064\\
        WRMF & 0.4229 & 0.3200 & 0.3783 & 0.2178 & 0.5762 & 0.4859 & 0.4923 & 0.3393 & 0.2745 & 0.1848 & 0.2623 & 0.1336\\
        $P^{3}\alpha$ & 0.4066 & 0.3086 & 0.3553 & 0.2016 & 0.5432 & 0.4539 & 0.4532 & 0.3028 & 0.2469 & 0.1635 & 0.2370 & 0.1154\\
        ItemKNN-$cos$ & 0.4088 & 0.3121 & 0.3577 & 0.2063 & 0.5599 & 0.4783 & 0.4664 & 0.3215 & 0.2683 & 0.1804 & 0.2566 & 0.1277\\
        SLIM & 0.4298 & 0.3249 & 0.3775 & 0.2147 & 0.5643 & 0.4710 & 0.4862 & 0.3284 & 0.2172 & 0.1343 & 0.2223 & 0.1008\\
        CAAE & 0.2217 & 0.1531 & 0.1941 & 0.0913 & 0.4767 & 0.3894 & 0.3902 & 0.2467 & 0.0834 & 0.0504 & 0.0940 & 0.0378\\
        CFGAN-u & 0.4044 & 0.3066 & 0.3487 & 0.1977 & 0.5123 & 0.4151 & 0.4224 & 0.2695 & 0.2358 & 0.1482 & 0.2338 & 0.1079\\
        CFGAN-i & 0.2211 & 0.1546 & 0.1909 & 0.0928 & 0.4462 & 0.3533 & 0.3707 & 0.2267 & 0.2219 & 0.1433 & 0.2145 & 0.1021\\
        GANMF-u & \textbf{0.4564} & \textbf{0.3551} & \textbf{0.4032} & \textbf{0.2423} & \underline{0.6076} & \underline{0.5255} & \underline{0.5151} & \underline{0.3715} & \underline{0.2857} & \underline{0.1936} & \textbf{0.2742} & \textbf{0.1402}\\
        GANMF-i & \underline{0.4500} & \underline{0.3503} & \underline{0.3985} & \underline{0.2399} & \textbf{0.6230} & \textbf{0.5445} & \textbf{0.5276} & \textbf{0.3866} & \textbf{0.2865} & \textbf{0.1943} & \underline{0.2725} & \underline{0.1397}\\
\bottomrule
    \end{tabular}
    \label{tab:baselines}
    \end{table*}
    
We note that we have omitted some GAN-based works. We have omitted IRGAN since CFGAN shows a clear improvement over it. $\text{RAGAN}^{\text{BT}}$ is intended to work with explicit ratings instead of implicit feeback. Morever, both $\text{RAGAN}^{\text{BT}}$ and AugCF use GAN to alleviate sparsity of URM and still rely entirely on traditional CF models to provide the recommendations. \cite{cai2018generative} and AugCF also rely on side information for their recommendations while we focus only on the user-item interactions.

\subsubsection{Discussion}
\label{sec:discussion}
We provide here a brief breakdown of the obtained results in table \ref{tab:baselines}. GANMF variants show superior performance against all baselines in the 3 datasets. In particular, GANMF performs on average 24\% better than CFGAN models despite using the same \emph{vector-wise} training procedure. We attribute this disparity to the autoencoder discriminator which is especially helpful when the generator is tasked to generate high dimensional vectors as in the case of RS. Additionally, between the two GANMF variants there is little difference which we were not expecting, especially for LastFM and MovieLens HetRec datasets where the ratio number of users to number of items is higher. On the other hand, CFGAN variants show the opposite behavior, with the user variant consistently surpassing the item one. For the other GAN-based algorithm, CAAE, we were not able to find an official implementation so we tried to implement it following only the paper. However, we found CAAE's recommendation quality to not match the one reported in the original paper. In our dataset splits it performed on par with the non-personalized technique.

The two latent factor models, PureSVD and WRMF have comparable performance between them in all datasets. The best performing GANMF variant is on average 15\% and 8\% better than PureSVD and WRMF, respectively, in the MovieLens datasets. In LastFM dataset, GANMF performs on average 29\% better than PureSVD. It is clear that the matrix factorization model learned by GANMF is able to find latent factors that explain better the interactions of users and items (we explore this in more details in section \ref{sec:mf_learned}). ItemKNN, SLIM and $P^{3}\alpha$ all model the item-item similarity matrix and as such, their performances do not differ much. They also tend to score slightly lower than MF techniques. Compared to GANMF, the best neighborhood model is 10\% worse across metrics and datasets. We conclude that GANMF is able to outperform baselines in GAN-based models and traditional approaches.

\subsection{Ablation Study}
\label{ablation_study}
In order to study the components of GANMF, we perform 2 different experiments where we replace one component at a time.

\subsubsection{GANMF with binary classifier discriminator}
\label{binganmf}
In this experiment we drop the autoencoder discriminator in GANMF and replace it with a binary classifier discriminator just like in vanilla cGAN and CFGAN. We denote this model \emph{binGANMF}. This new discriminator outputs the probability of its input coming from the URM. In order to evaluate only the impact of the discriminator, we keep everything else exactly as explained in section \ref{ganmf}; generator with embedding layers and feature matching. As user features to match, we use the learned features in the last layer of the discriminator before the final output. We retune again binGANMF with bayesian optimization. On table \ref{tab:ablation_study} we report its performance on the test set along with the standard GANMF. We see that when equipped with an autoencoder discriminator, GANMF variants performs much better than binGANMF on all metrics, as much as 4 times better on LastFM dataset on MAP@20 metric.

\begin{table*}[htbp]
    \centering
    \caption{Ablation study. The best results per dataset and per metric are given in bold, the second best results are underlined.}
    \normalsize
    \begin{tabular}{c|cccc|cccc|cccc}
        \toprule
        \multirow{3}{*}{Algorithm}
        & \multicolumn{4}{|c}{\textbf{ML 1M}} & \multicolumn{4}{|c}{\textbf{ML HetRec}} & \multicolumn{4}{|c}{\textbf{LastFM}}\\
        & \textbf{NDCG} & \textbf{MAP} & \textbf{NDCG} & \textbf{MAP} & \textbf{NDCG} & \textbf{MAP} & \textbf{NDCG} & \textbf{MAP} & \textbf{NDCG} & \textbf{MAP} & \textbf{NDCG} & \textbf{MAP}\\
        & @5 & @5 & @20 & @20 & @5 & @5 & @20 & @20 & @5 & @5 & @20 & @20\\
        \midrule
        GANMF-u & \textbf{0.4564} & \textbf{0.3551} & \textbf{0.4032} & \textbf{0.2423} & \underline{0.6076} & \underline{0.5255} & \underline{0.5151} & \underline{0.3715} & \underline{0.2857} & \underline{0.1936} & \textbf{0.2742} & \textbf{0.1402}\\
        GANMF-i & \underline{0.4500} & \underline{0.3503} & \underline{0.3985} & \underline{0.2399} & \textbf{0.6230} & \textbf{0.5445} & \textbf{0.5276} & \textbf{0.3866} & \textbf{0.2865} & \textbf{0.1943} & \underline{0.2725} & \underline{0.1397}\\
        binGANMF-u & 0.2185 & 0.1480 & 0.1920 & 0.0895 & 0.4660 & 0.3776 & 0.3861 & 0.2426 & 0.0790 & 0.0454 & 0.0842 & 0.0337\\
        binGANMF-i & 0.2889 & 0.2075 & 0.2446 & 0.1240 & 0.4707 & 0.3827 & 0.3871 & 0.2440 & 0.0855 & 0.0477 & 0.0966 & 0.0374\\
        \bottomrule
    \end{tabular}
    \label{tab:ablation_study}
\end{table*}

\begin{figure*}[htbp]
    \centering
    \subfigure[ML 1M]{{\includegraphics[width=0.32\linewidth]{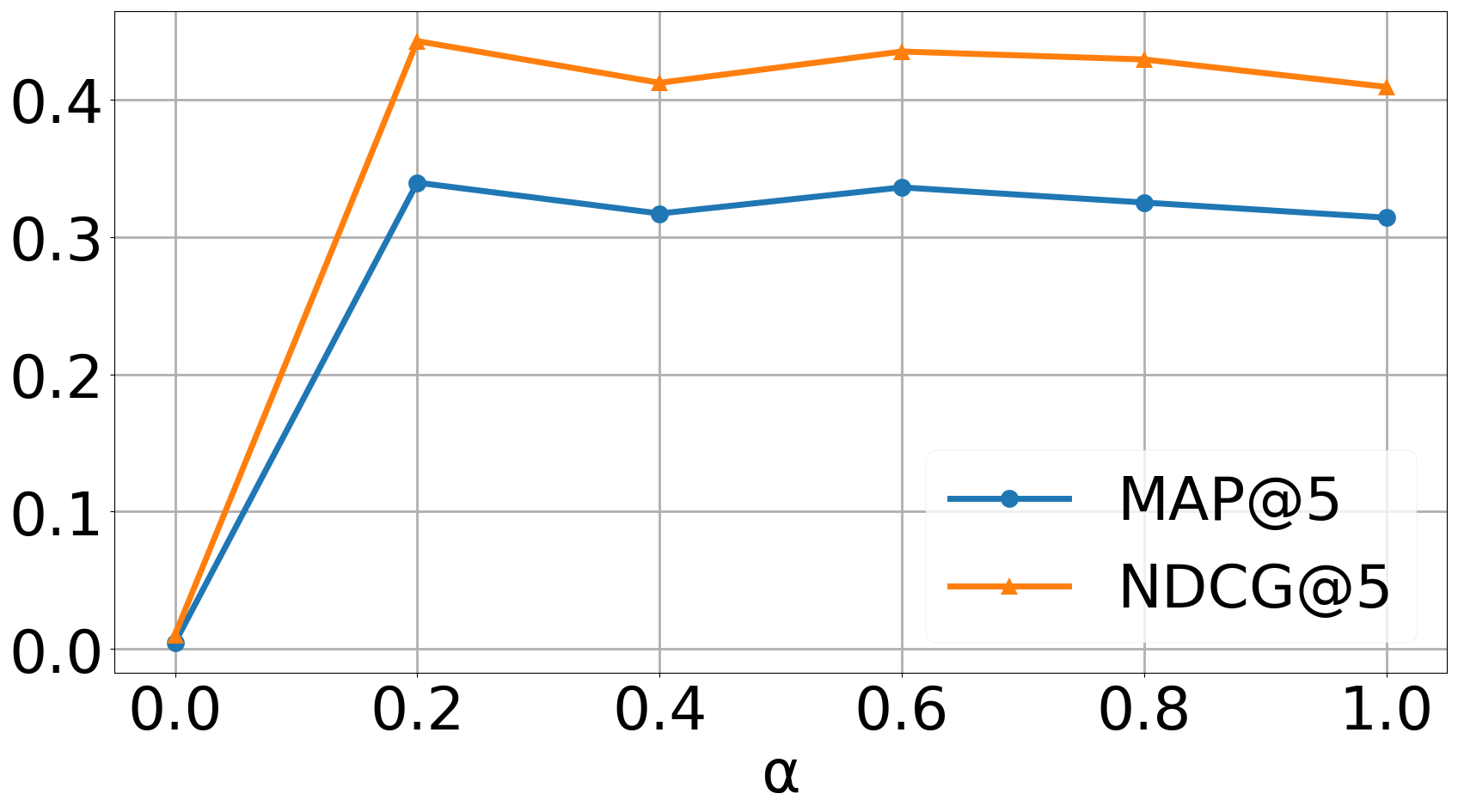}}}\hfill
    \subfigure[ML HetRec]{{\includegraphics[width=0.32\linewidth]{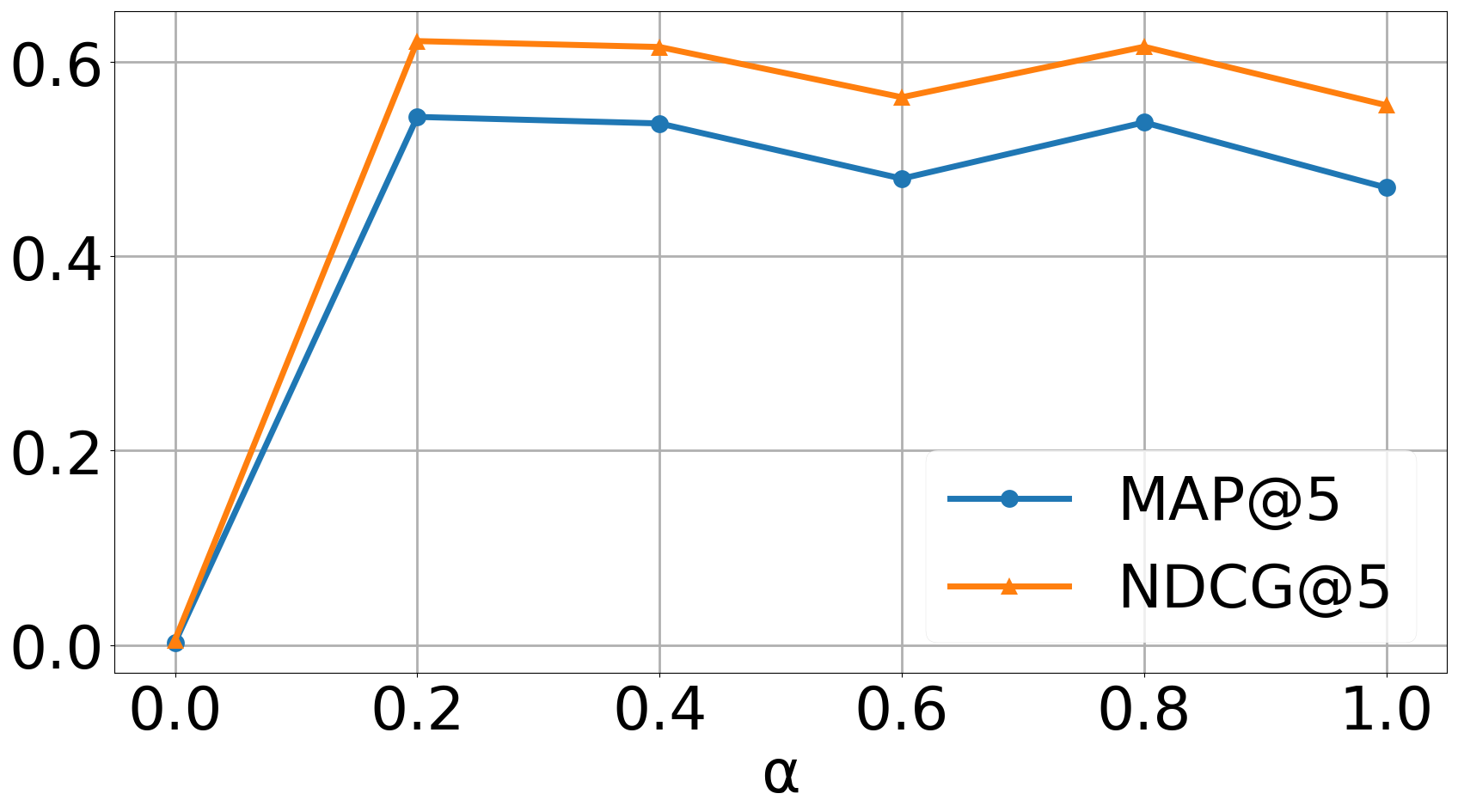}}}\hfill
    \subfigure[LastFM]{{\includegraphics[width=0.32\linewidth]{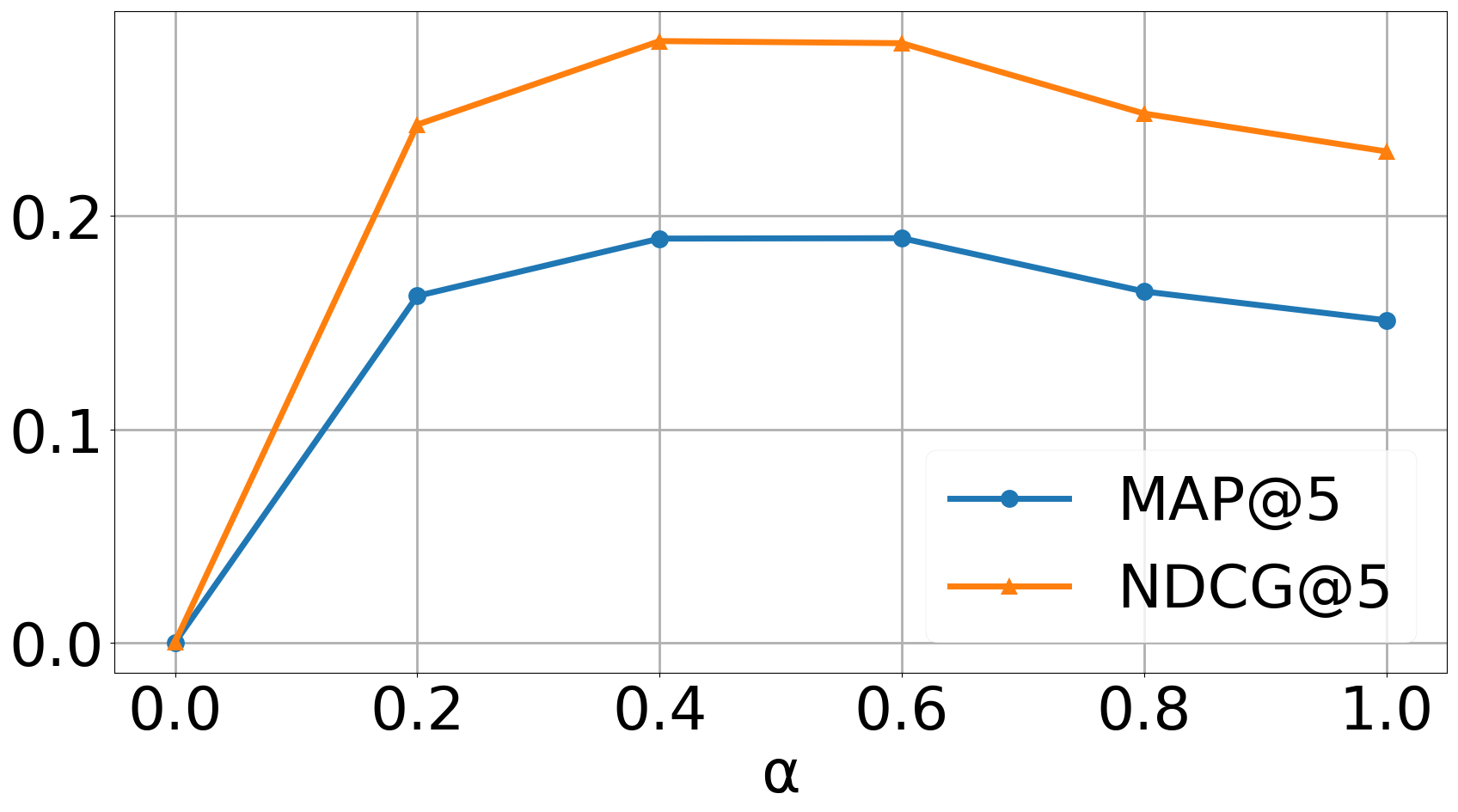}}}
    \caption{Effect of feature matching loss on GANMF-i performance on MovieLens 1M, MovieLens HetRec and LastFM.}
    \label{fig:fm_all}
\end{figure*}

\subsubsection{Effect of Feature Matching}
\label{feature_matching}
In order to understand how feature matching affects GANMF, we modify only the feature matching coefficient $\alpha$ in equation \ref{eq:ganloss_fm} in the range $[0-1]$ with a step of 0.2 and rerun bayesian optimization again for the resulting GANMF variants. In figure \ref{fig:fm_all} we show how different values of $\alpha$ change the performance of GANMF for MovieLens 1M, MovieLens HetRec and LastFM, respectively (for space reasons we give only GANMF-i, similar behavior is observed for GANMF-u). On all datasets and the 2 different cutoffs, a combination of both the adversarial GAN loss and feature matching provides the best results for GANMF.

\begin{figure*}[!hbtp]
    \centering
    \subfigure[ML 1M w/o feature matching. Mean: 0.9999, Std: 0.000004]{{\includegraphics[width=0.32\linewidth]{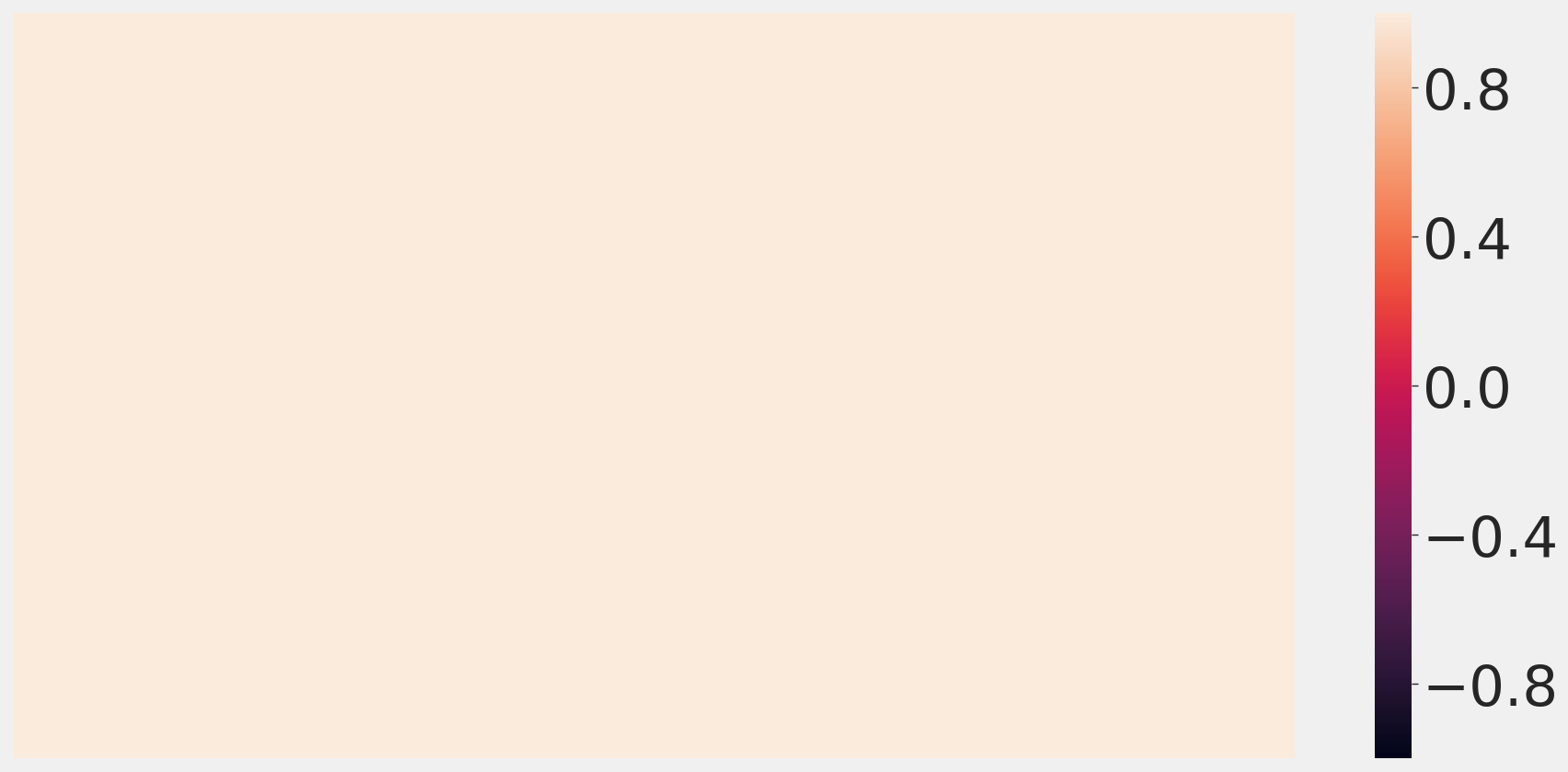}}}\hfill
    \subfigure[LastFM w/o feature matching. Mean: 0.9989, Std: 0.0018.]{{\includegraphics[width=0.32\linewidth]{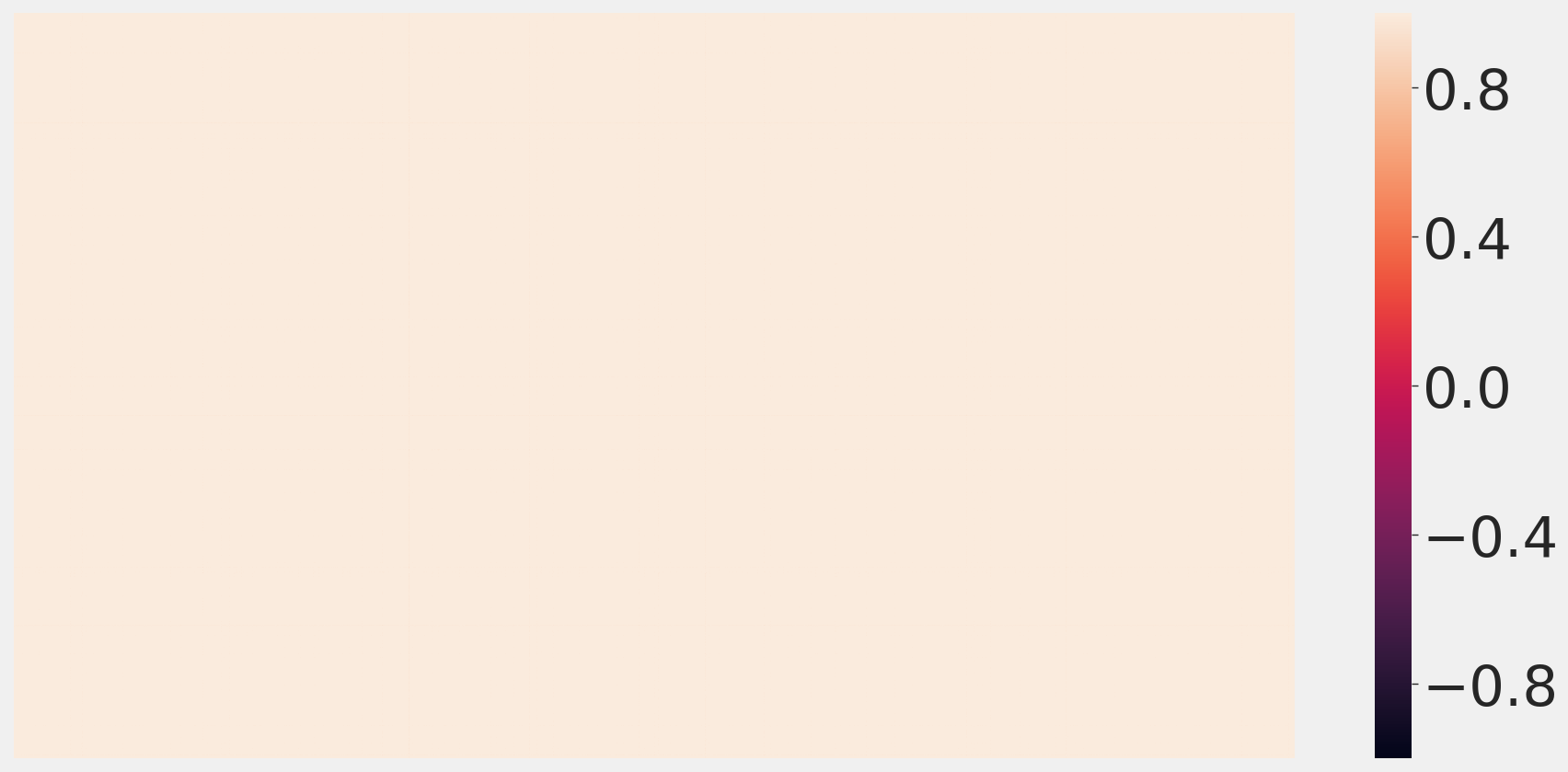}}}\hfill
    \subfigure[ML HetRec w/o feature matching. Mean: 0.9999, Std: 0.000006.]{{\includegraphics[width=0.32\linewidth]{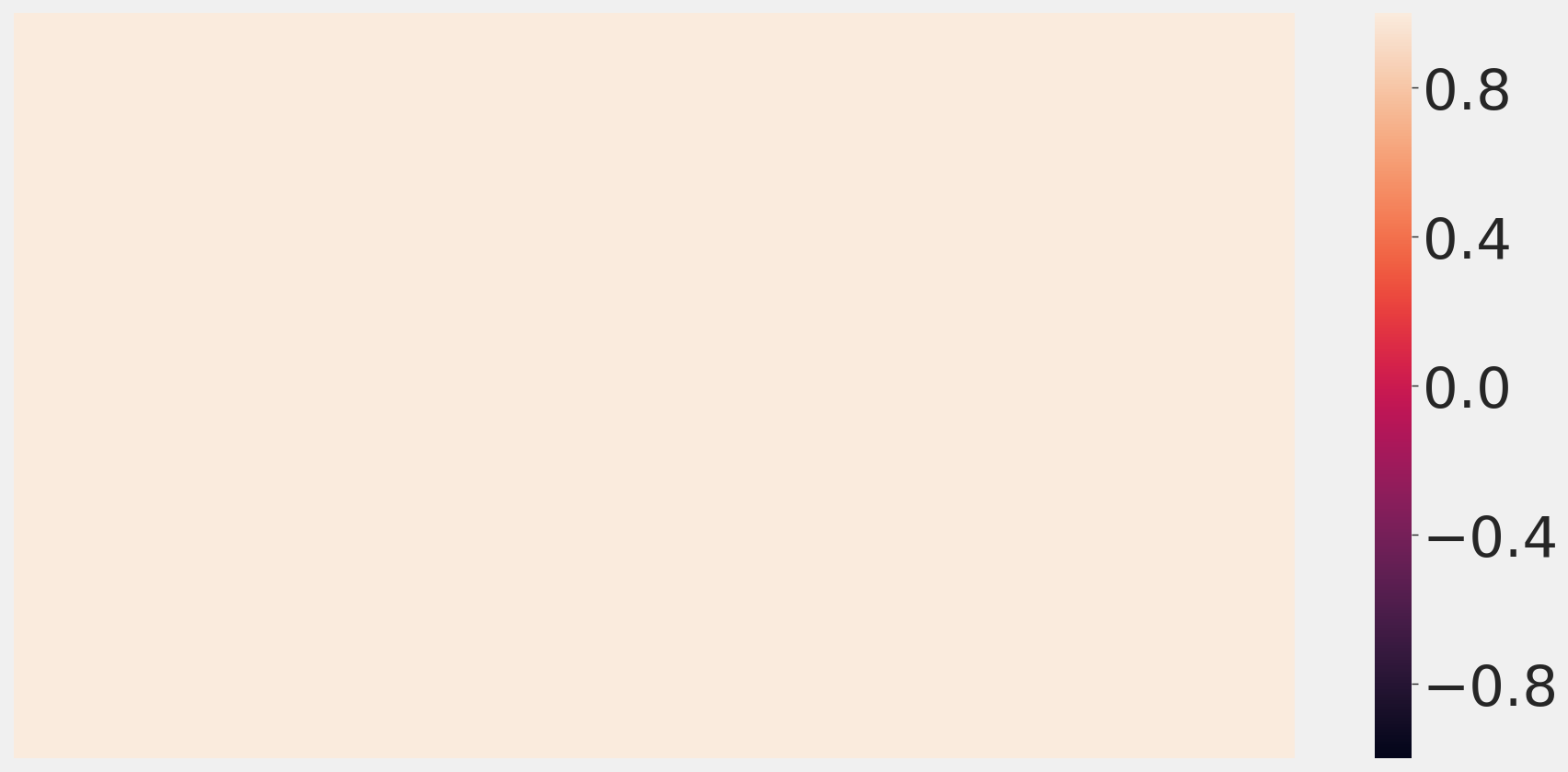}}}\\
    \subfigure[ML 1M w/ feature matching. Mean: 0.3644, Std: 0.1952]{{\includegraphics[width=0.32\linewidth]{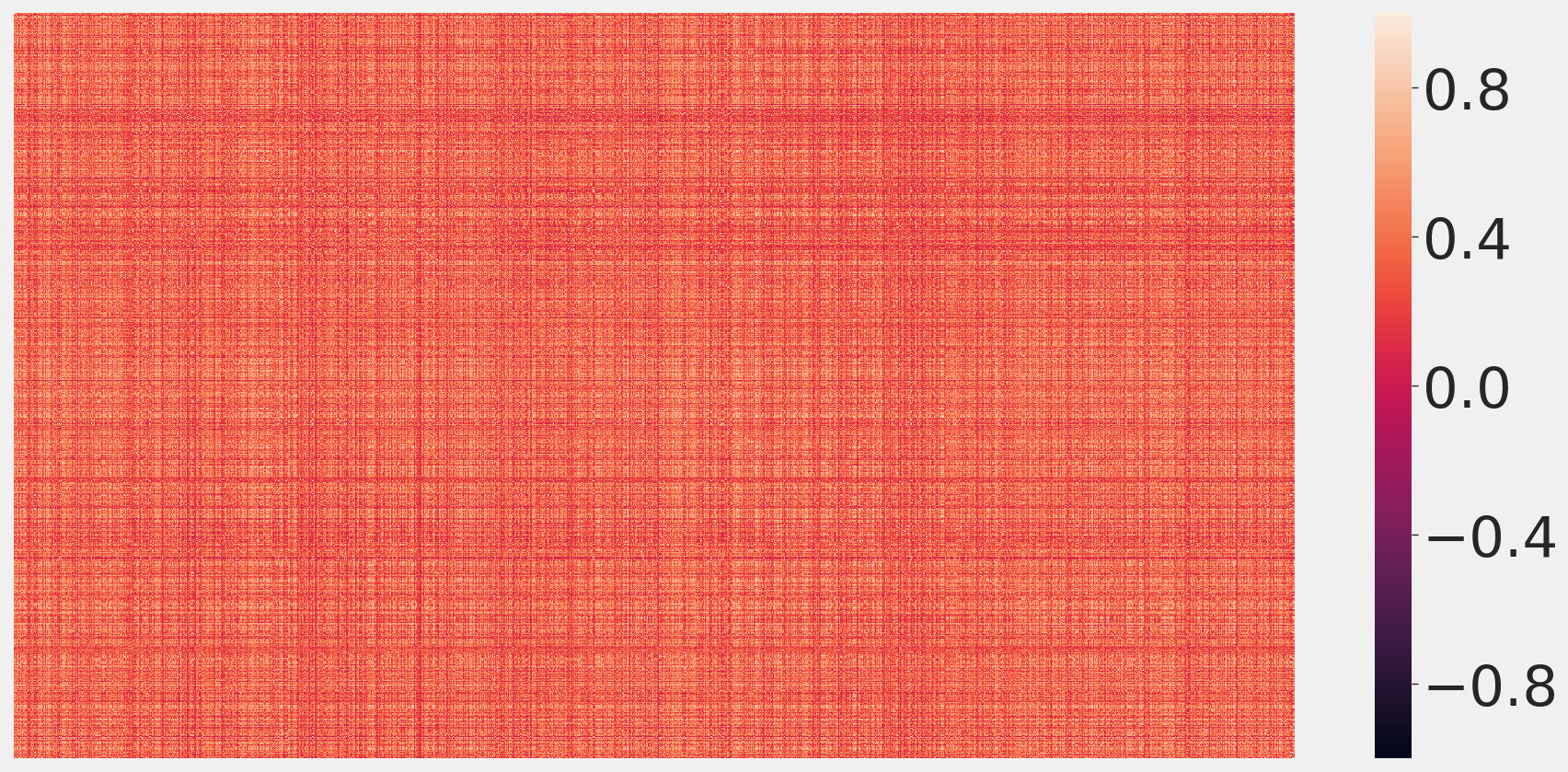}}}\hfill
    \subfigure[LastFM w/ feature matching. Mean: 0.1535, Std: 0.2161.]{{\includegraphics[width=0.32\linewidth]{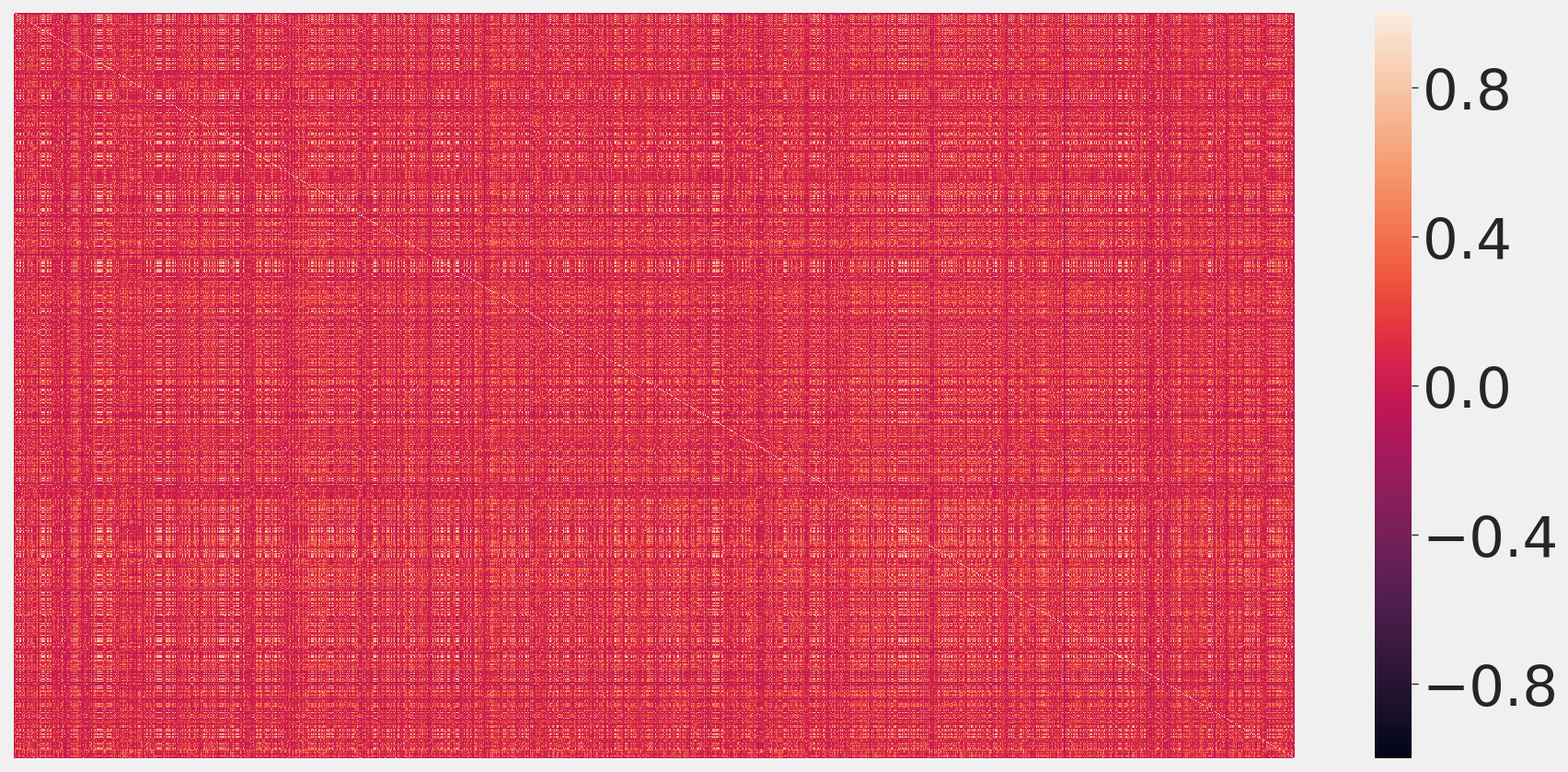}}}\hfill
    \subfigure[ML HetRec w/ feature matching. Mean: 0.5597, Std: 0.2523.]{{\includegraphics[width=0.32\linewidth]{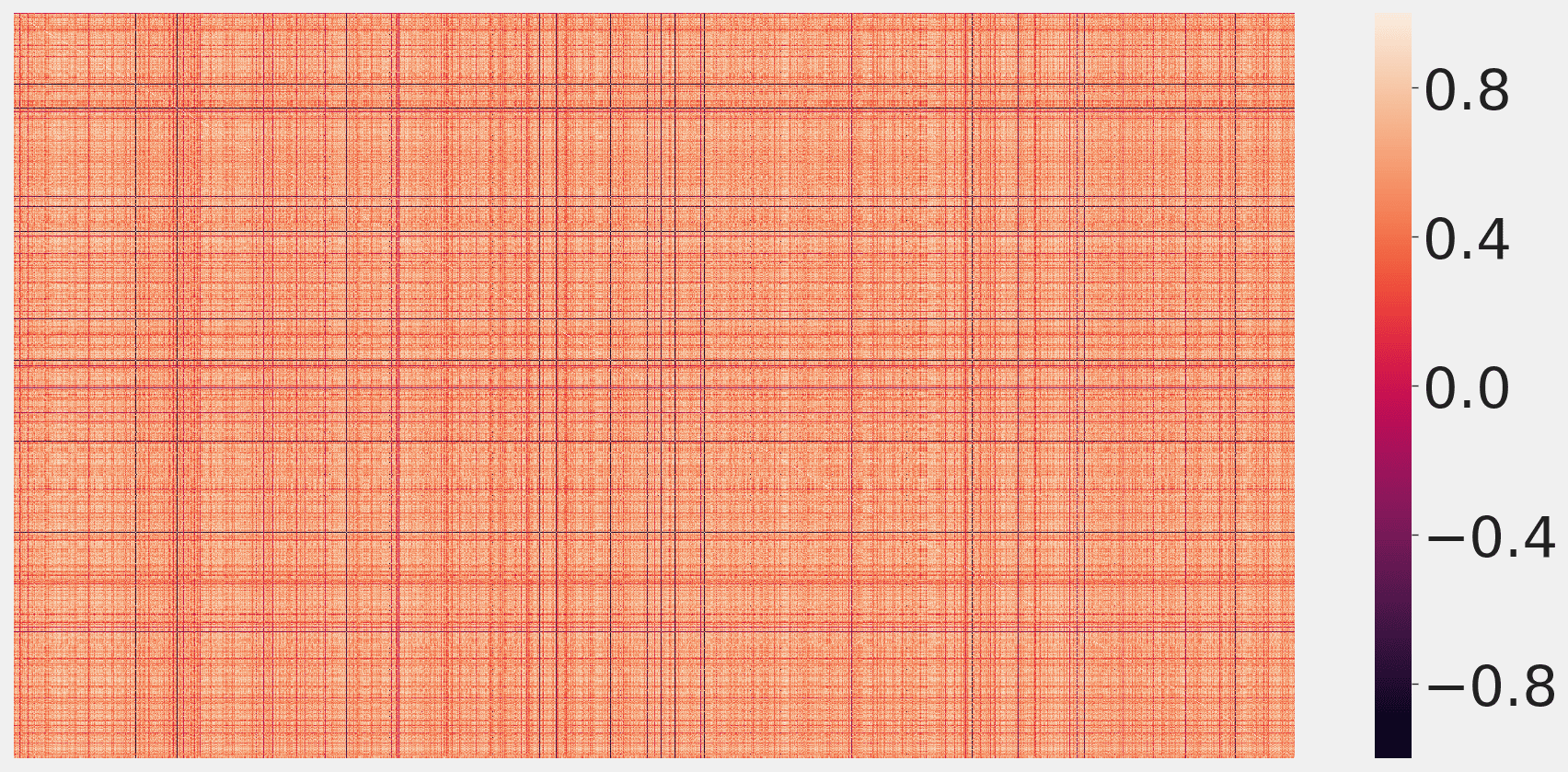}}}
    \caption{Feature matching conditioning on the user generated profiles by GANMF-u. The heatmaps represent the user-user similarity with and without feature matching. Mean and standard deviation of the similarities are given for each dataset.}
    \label{fig:fm_conditioning}
\end{figure*}

The other important aspect of using feature matching is to enforce conditional generation for the generator. To investigate its usefulness, we train GANMF with and without feature matching and after each training phase, we produce profiles for all users. Then, we compute the cosine similarity between each pair of users and show them as heatmaps in figure \ref{fig:fm_conditioning}. We observe significant decrease in user-user similarity after the application of feature matching. This means that the generator outputs less similar vectors and each user conditioning attribute is mapped to a more unique profile.

\subsection{On the MF model learned by GANMF}
\label{sec:mf_learned}
Given the quantitative advantage of GANMF over the considered baselines, it is important to investigate the latent factors model learned by GANMF.

\begin{figure*}[htbp]
    \centering
    \subfigure[ML 1M]{{\includegraphics[width=0.32\linewidth]{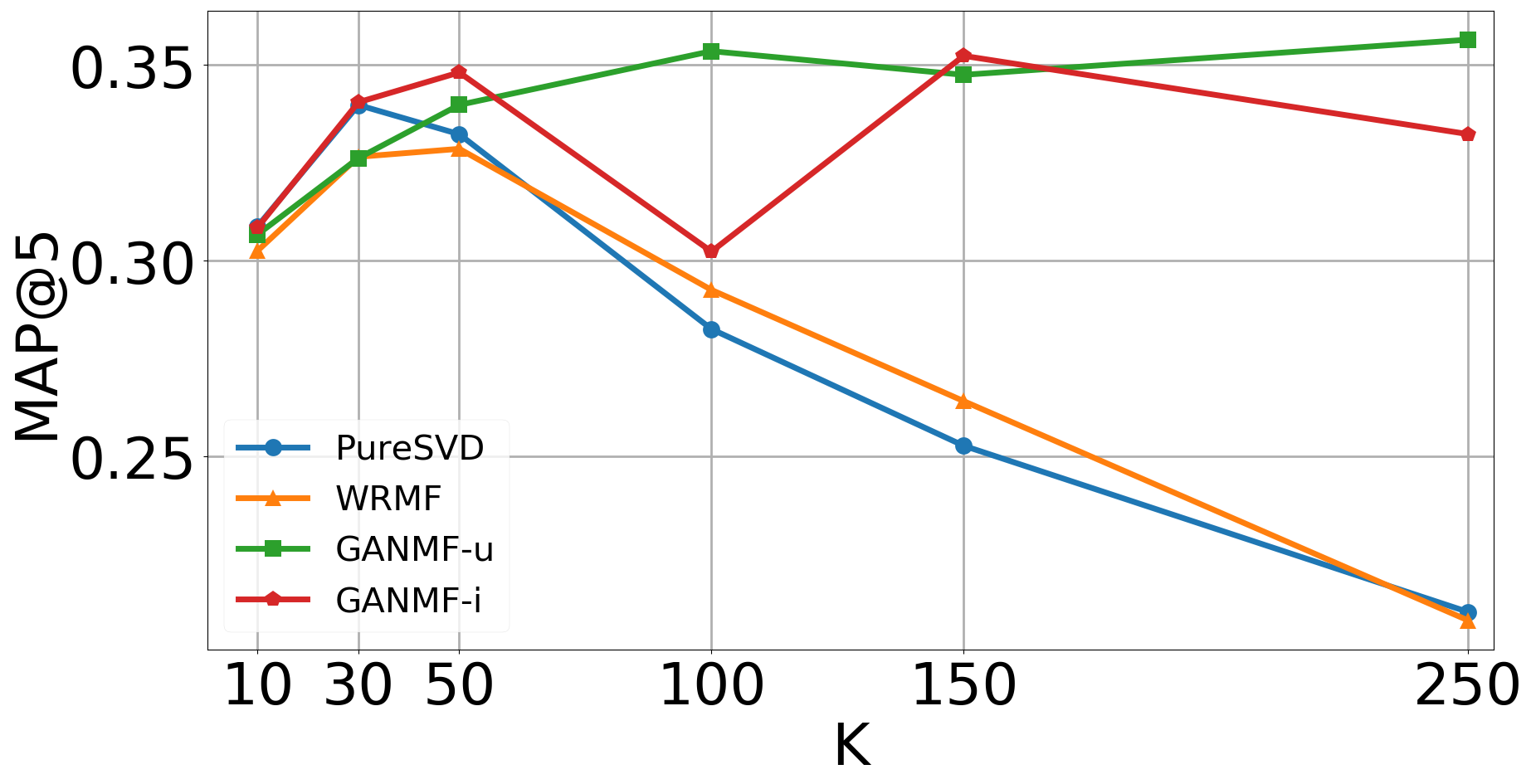}}}\hfill
    \subfigure[ML HetRec]{{\includegraphics[width=0.32\linewidth]{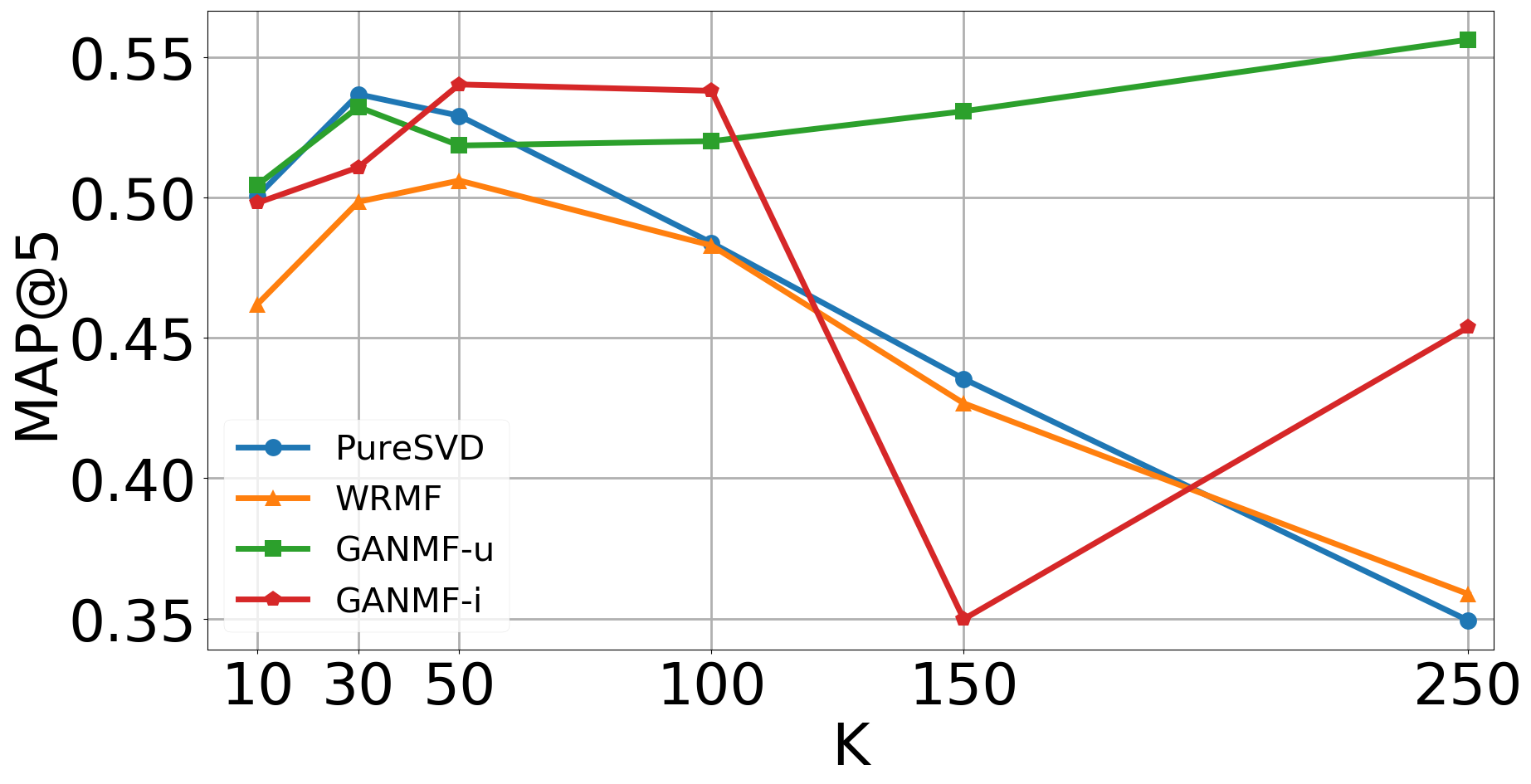}}}\hfill
    \subfigure[LastFM]{{\includegraphics[width=0.32\linewidth]{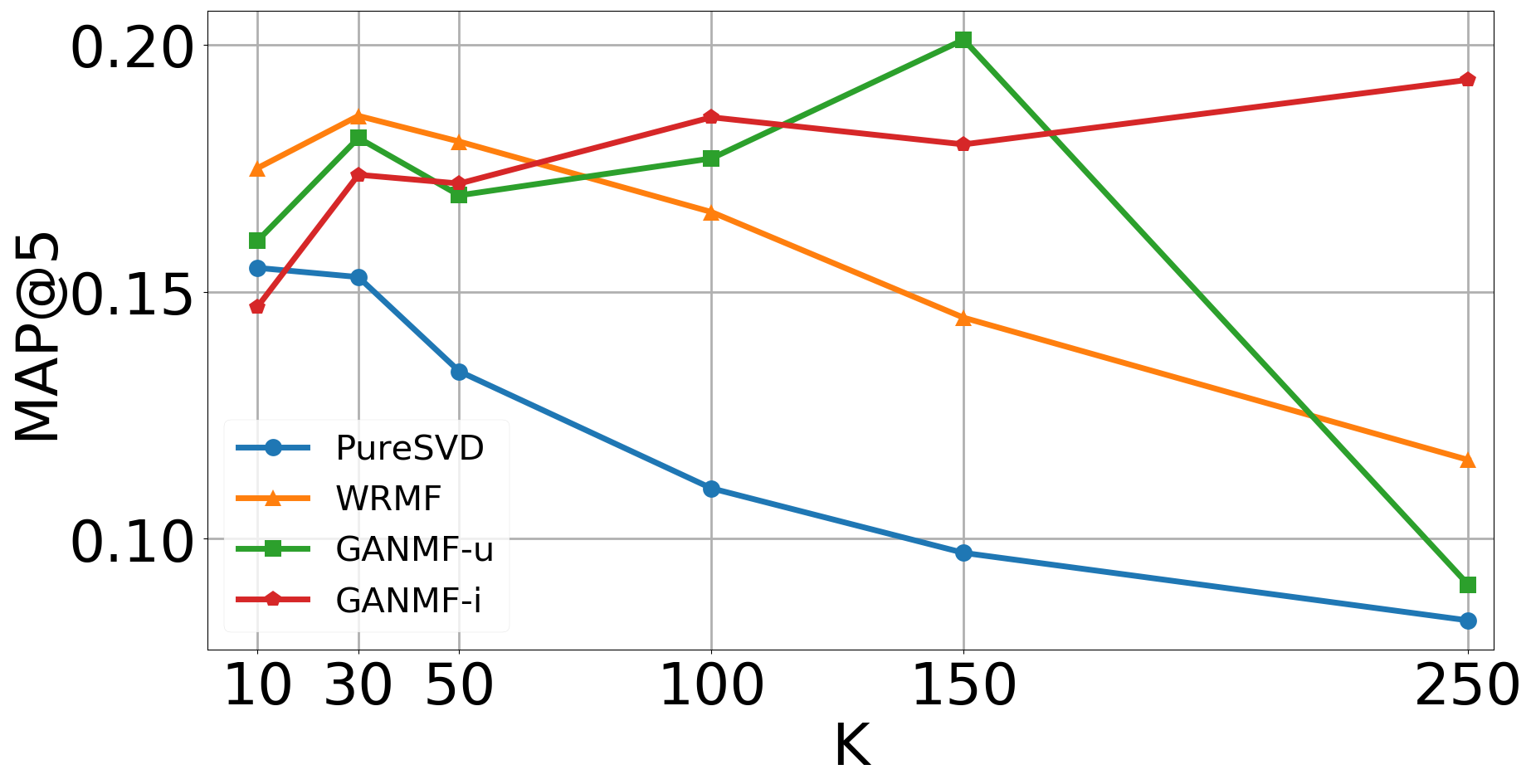}}}
    \caption{MAP@5 of PureSVD, WRMF and GANMF on MovieLens 1M, MovieLens HetRec and LastFM for varying latent factors number.}
    \label{fig:lf_all}
\end{figure*}

We compare the behavior of GANMF with varying number of latent factors $K$ along with two other MF baselines, PureSVD and WRMF. For each model we fixed $K$ \textbf{and tuned the other hyperparameters} following the procedure detailed in section \ref{settings_evaluation} with 35 runs instead of 50. Their performance on MAP@5 is given in figure \ref{fig:lf_all}. GANMF dominates the other baselines in most of the considered latent factors, with the exception of $K < 100$ on LastFM where WRMF performs better. The best models for PureSVD and WRMF tend to provide more accurate recommendations when using a relatively low number of latent factors since such traditional MF techniques are known to overfit with large $K$ \cite{johnson2014logistic}. GANMF on the other hand still improves its recommendation accuracy with increasing $K$. This is mainly due to GANMF using the discriminator to learn the latent factors instead of learning them directly from the training data. This is further reinforced by the fact that we do not place a regularization term on the parameters of $\mathcal{G}$.

\begin{figure*}[htbp]
    \centering
    \subfigure[ML 1M]{{\includegraphics[width=0.32\linewidth]{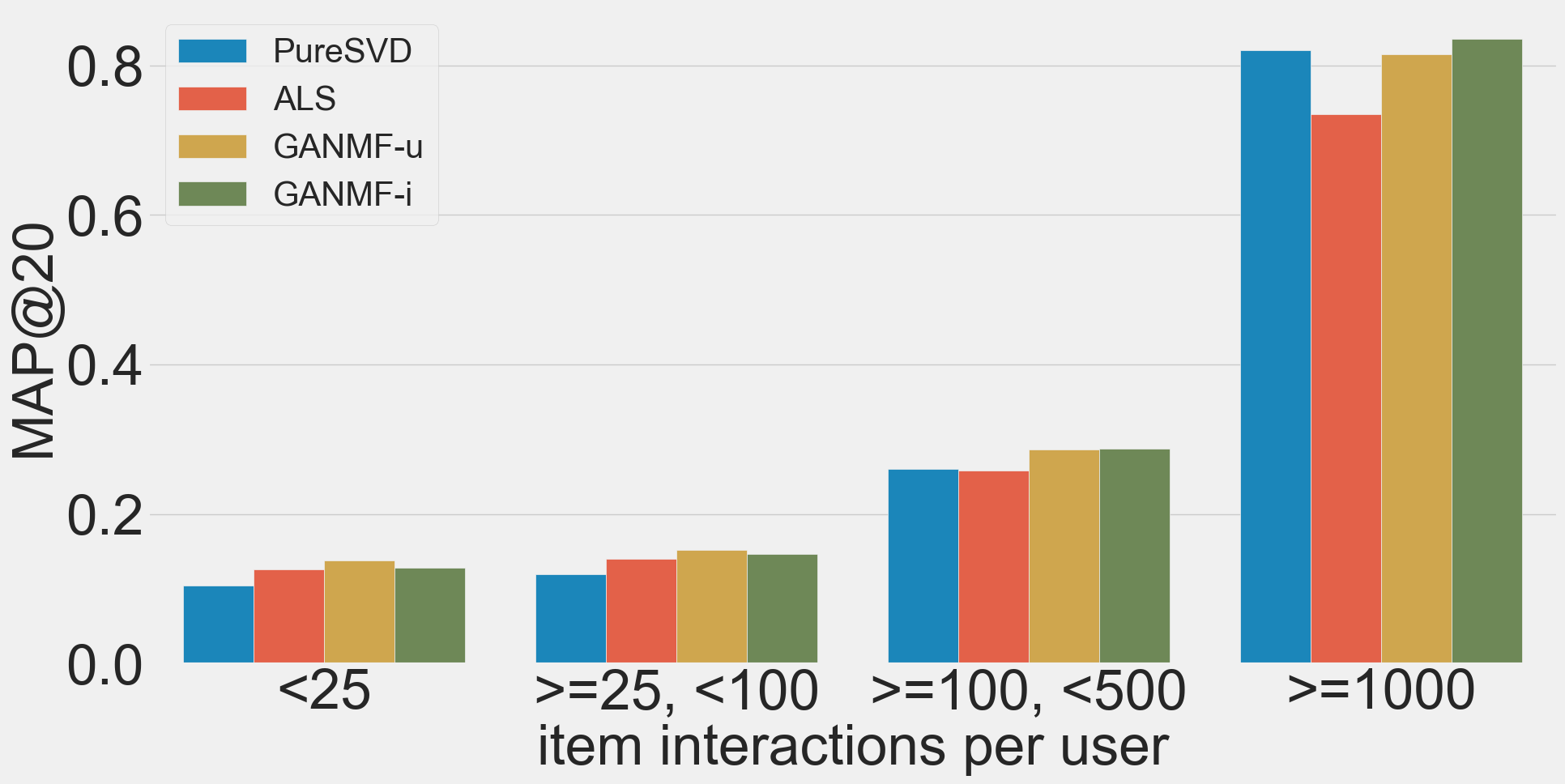}}}\hfill
    \subfigure[ML HetRec]{{\includegraphics[width=0.32\linewidth]{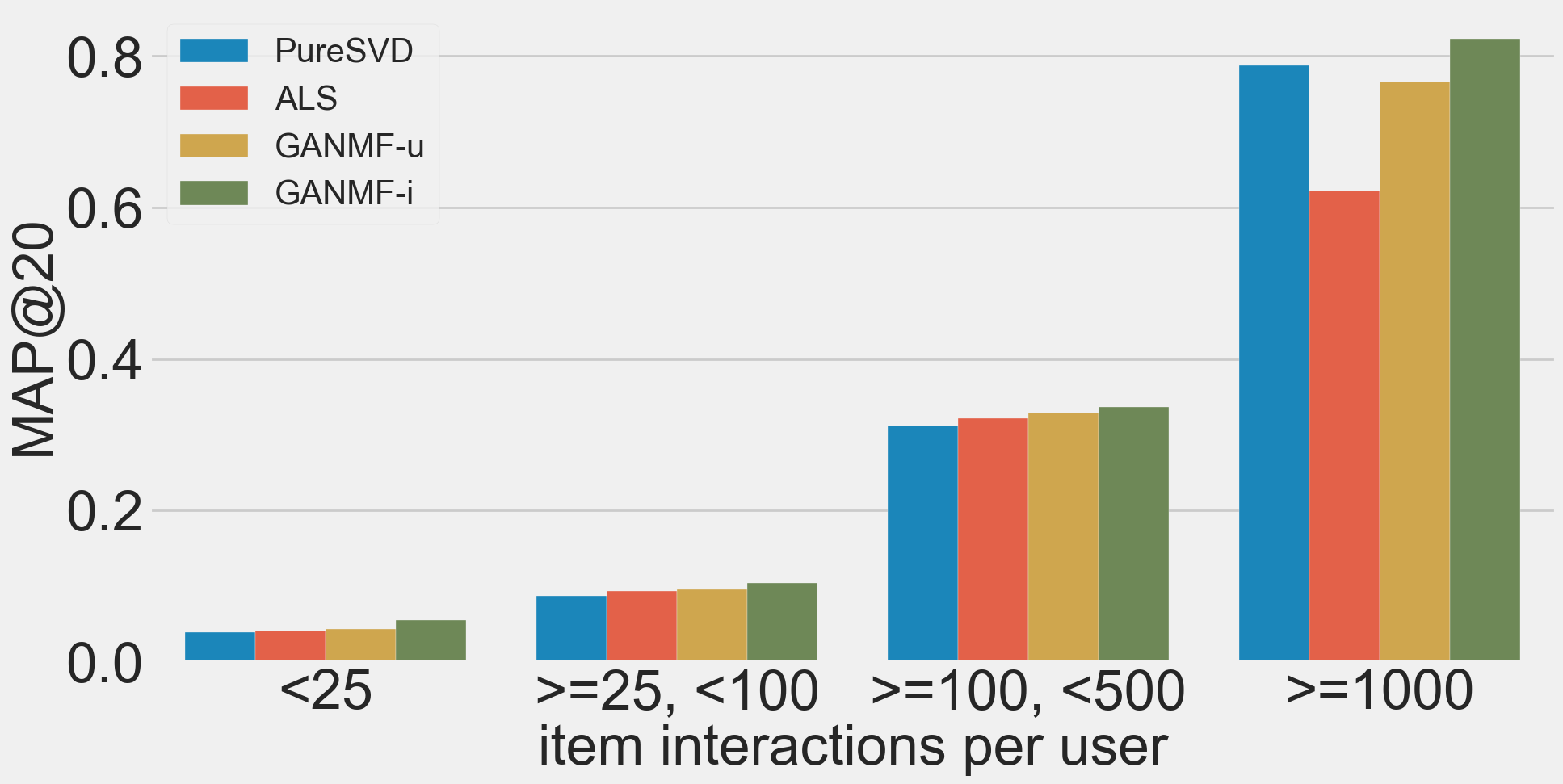}}}\hfill
    \subfigure[LastFM]{{\includegraphics[width=0.32\linewidth]{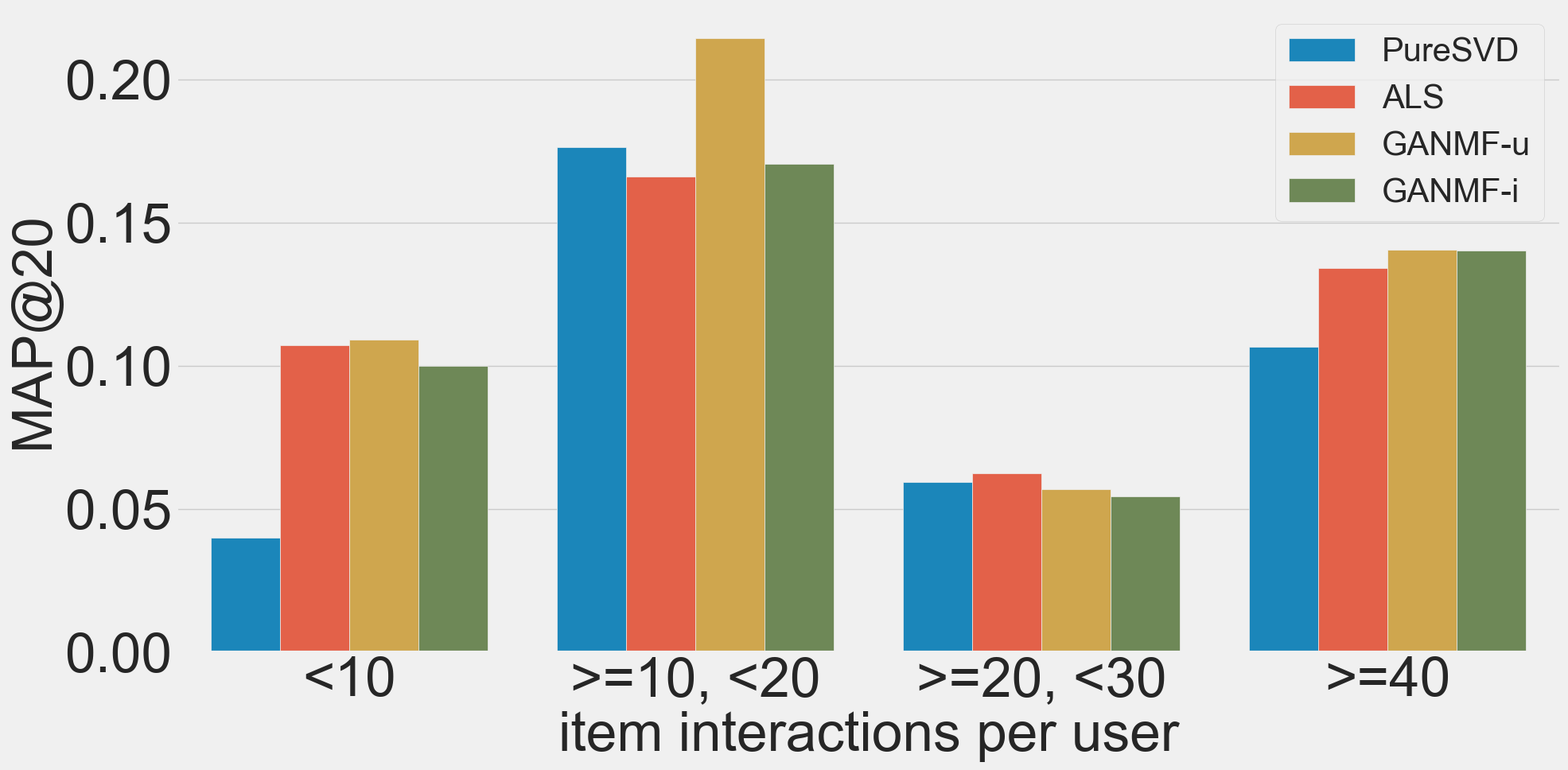}}}
    \caption{MAP@20 of PureSVD, WRMF and GANMF on MovieLens 1M, MovieLens HetRec and LastFM for varying latent factors number.}
    \label{fig:user_types_all}
\end{figure*}

Another important aspect is the behavior of GANMF on the different types of users based on the number of items they have interacted with. Figure \ref{fig:user_types_all} shows the performance of PureSVD, WRMF and GANMF for 4 types of users. MF models are known to be susceptible to the \emph{cold start} problem with performance suffering for users that have interacted with few items \cite{park2009pairwise}. This is evident especially on MovieLens 1M and MovieLens HetRec where ranking accuracy is much lower for users that have interacted with less than 25 items and less than 100 items. Even for these users GANMF is able to provide better recommendations than the MF baselines. This can be attributed to the autoencoder discriminator and feature matching loss. In its coding layer, the discriminator $\mathcal{D}$ learns a meaningful representations of users' preference over the items. Thus, even users that do not share exactly the same items in data space, might share similar latent representation in the coding layer. When trained with feature matching, the generator can transfer some knowledge from the shared latent space back to the latent factors of users with few interactions.

\section{Conclusion}
\label{conclusion}
In this work we presented GANMF, a novel approach to building GAN-based latent factors models for top-N recommendation with implicit feedback. We identified 2 main issues when using cGAN in CF; using a single-output binary-classifier discriminator does not provide rich enough gradients to the generator given the dimensionality of the user/item profiles in RS and the lacking of multiple data samples per user which causes the generator of a cGAN to disregard the conditioning attribute in the case of CF. We give solutions for both of these issues by replacing the binary-classifier discriminator in the original formulation of cGAN with an autoencoder and by incorporating a feature matching loss in the generator. Through an ablation study we show that our model can achieve its best performance with a combination of both GAN loss and feature matching loss. More importantly, we show that feature matching enforces conditional generation of user/item profiles. We tested our proposed model in 3 publicly available datasets of different sparsity and shapes and provided a comparison with 8 other baselines, representatives for latent factor, neighborhood and GAN-based models. We showed that GANMF outperformed all of them in 2 ranking metrics across 3 datasets. Finally, the qualitative results we provided on the latent factors model learned by GANMF indicate its efficiency for users with few interactions. These results show that GAN are a promising technique that can be used as a main component in RS.

\bibliographystyle{ACM-Reference-Format}
\bibliography{sample-bibliography} 

\end{document}